%% file: ANA-STDM-2019-28-PAPER.tex
\documentclass[cernpreprint, texlive=2025, atlasdraft=false, UKenglish, texmf, orcidlogo]{atlasdoc}
\usepackage[subcaption]{atlaspackage}
\usepackage{atlasbiblatex}

\usepackage[unit=false, hepparticle=true, hepprocess=true, jetetmiss=true, misc=true, process=true]{atlasphysics}

\usepackage{atlaspackage}
\usepackage{atlasphysics}
\usepackage{atlasbiblatex}

\graphicspath{{logos/}{figures/}}

\usepackage{multirow}
\usepackage{adjustbox}
\usepackage{physics}
\usepackage[compat=1.1.0]{tikz-feynman}
\usepackage{ANA-STDM-2019-28-PAPER-defs}

\AtlasTitle{
Observation of $W^{+}W^{-}\gamma$ production in $pp$ collisions at $\sqrt{s}$ = 13 TeV with the ATLAS detector and constraints on anomalous quartic gauge-boson couplings
}

\AtlasAbstract{%
This Letter reports the observation of $W^{+}W^{-}\gamma$ triboson production in 140 fb$^{-1}$ of data collected by the ATLAS detector from proton--proton collisions at a centre-of-mass energy of $\sqrt{s}$ = 13 TeV at the LHC. Events with an opposite-charge $e\mu$ pair, a high transverse-momentum photon, and significant missing transverse momentum are considered. The observed (expected) significance of the signal is 5.9 (6.0) standard deviations. The measured fiducial cross-section, defined for the $W^{+}W^{-}\gamma\to e^{\pm}\mu^{\mp}\nu\bar{\nu}\gamma$ final state is 6.2 $\pm$ 0.8 (stat.) $\pm$ 0.6 (sys.) fb, in good agreement with the Standard Model prediction of 6.1$^{\,+1.0}_{-0.7}$ fb. Constraints on the Wilson coefficients of 13 dimension-8 operators describing physics beyond the Standard Model through anomalous quartic gauge-boson couplings are derived using the effective field theory framework.}

\AtlasRefCode{STDM-2019-28}
\PreprintIdNumber{CERN-EP-2025-187}

\AtlasJournal{Phys.\ Lett.\ B.}

\AtlasJournalRef{\PLB 873 (2026) 140050}
\AtlasDOI{10.1016/j.physletb.2025.140050}

\hypersetup{pdftitle={ATLAS document},pdfauthor={The ATLAS Collaboration}}

\begin{document}

\maketitle

\clearpage
\section{Introduction}
%


In the Standard Model (SM), the electroweak interaction is based on the non-Abelian SU(2)$_\text{L}\times$ U(1)$_\text{Y}$ gauge group. This leads to couplings among the electroweak gauge bosons (the \PW and \PZ bosons and the photon) through triple and quartic vertices~\cite{Godfrey:1995cd}. Measurements of triboson production processes allow stringent tests of the gauge structure and offer sensitivity to physics beyond the SM.

The ATLAS~\cite{PERF-2007-01} and CMS~\cite{CMS-CMS-00-001} collaborations at the Large Hadron Collider (LHC)~\cite{Bruning:782076} investigated both the \wwy and \wzy processes~\cite{STDM-2016-05,CMS-SMP-13-009} using datasets collected at a centre-of-mass energy ($\sqrt{s}$) of \SI{8}{\TeV} and integrated luminosities ($\mathcal{L}$) of $\SI{20.2}{\ifb}$ and $\SI{19.7}{\ifb}$, respectively.
The CMS Collaboration has since observed the \wwy process at $\sqrt{s} = \SI{13}{\TeV}$ using its full Run-2 dataset corresponding to $\mathcal{L} = \SI{138}{\ifb}$~\cite{CMS-SMP-22-006}.
Many triboson production processes have been studied by both the ATLAS and CMS collaborations using their full Run-2 datasets, including $WWW$~\cite{STDM-2019-09,CMS-SMP-19-014}, $W\gamma\gamma$~\cite{STDM-2018-33,CMS-SMP-19-013},
$Z\gamma\gamma$~\cite{STDM-2021-09,CMS-SMP-19-013}, $WZ\gamma$~\cite{STDM-2019-17,CMS-SMP-22-018},
and $WVZ$, where $V$ refers to \PW or \PZ bosons~\cite{STDM-2020-08,CMS-SMP-19-014}. Both the ATLAS and CMS collaborations have set constraints on anomalous quartic gauge-boson coupling (aQGC) parameters.

\begin{figure}[htbp]
\centering

\begin{subfigure}[b]{0.25\textwidth}%
\includegraphics[width=\textwidth]{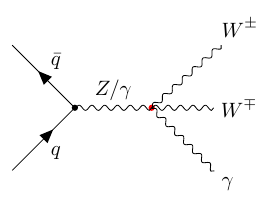}
\caption{}
\end{subfigure}
\begin{subfigure}[b]{0.20\textwidth}%
\includegraphics[width=\textwidth]{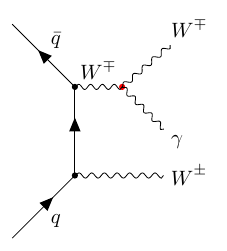}
\caption{}
\end{subfigure}%
\begin{subfigure}[b]{0.20\textwidth}%
\includegraphics[width=\textwidth]{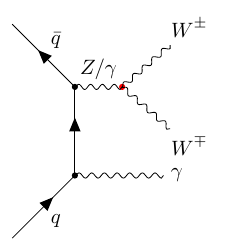}
\caption{}
\end{subfigure}
\begin{subfigure}[b]{0.20\textwidth}%
\includegraphics[width=\textwidth]{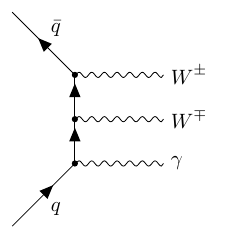}
\caption{}
\end{subfigure}%

\caption{
Representative leading-order Feynman diagrams for the \wwy production process with a triple or quartic gauge-boson coupling vertex shown in red:\
(a) diagram featuring a QGC vertex
(b,c) diagrams with a TGC vertex, and
(d) a diagram showing photon radiation from an initial-state quark.
Additional contributions arise from other diagrams not shown here.}

\label{fig:feynman_diagrams}
\end{figure}

This Letter reports the observation of \wwy production by examining the \emuy final state in \lumi~\cite{DAPR-2021-01} of data from proton--proton collisions at $\sqrt{s}=\SI{13}{\TeV}$ collected by the ATLAS detector at the LHC.
Events with an opposite-charge $e\mu$ pair, a photon, and missing transverse momentum are selected. The photon in a \wwy event can originate from three processes:\ direct coupling to a $\PW$ boson -- either triple gauge-boson coupling (TGC) or quartic gauge-boson coupling (QGC); initial-state radiation (ISR) from an initial quark; or emission as final-state radiation (FSR). In the case of FSR, the photon is emitted either by a lepton from a $\PW$ boson decay or by an associated quark.
\Cref{fig:feynman_diagrams} shows examples of the lowest-order Feynman diagrams for \wwy production where the photon arises from a QGC vertex, a TGC vertex, or is radiated directly from a quark without involving interactions between electroweak force carriers.

The fiducial cross-section is measured using a binned maximum-likelihood fit performed on the output distribution of a boosted decision tree, which was trained to enhance the separation between signal and background. Limits on aQGCs are also derived by performing an analysis in which deviations from the SM predictions are parameterised using effective field theory (EFT)~\cite{Eboli_EFT}.


%

%
\section{ATLAS detector}
%

\newcommand{\AtlasCoordFootnote}{%
ATLAS uses a right-handed coordinate system with its origin at the nominal interaction point (IP)
in the centre of the detector and the \(z\)-axis along the beam pipe.
The \(x\)-axis points from the IP to the centre of the LHC ring,
and the \(y\)-axis points upwards.
Cylindrical coordinates \((r,\phi)\) are used in the transverse plane,
\(\phi\) being the azimuthal angle around the \(z\)-axis.
The pseudorapidity is defined in terms of the polar angle \(\theta\) as \(\eta = -\ln \tan(\theta/2)\).
Angular distance is measured in units of \(\Delta R \equiv \sqrt{(\Delta\eta)^{2} + (\Delta\phi)^{2}}\).}

The ATLAS experiment~\cite{PERF-2007-01} at the LHC is a multipurpose particle detector with a forward--backward symmetric cylindrical geometry and nearly \(4\pi\) coverage in solid angle.\footnote{\AtlasCoordFootnote}
It consists of an inner tracking detector (ID) surrounded by a thin superconducting solenoid providing a \SI{2}{\tesla} axial magnetic field, electromagnetic and hadron calorimeters, and a muon spectrometer (MS).
The ID covers the pseudorapidity range \(|\eta| < 2.5\).
It consists of silicon pixel, silicon microstrip, and transition radiation tracking detectors.
Lead/liquid-argon (LAr) sampling calorimeters provide electromagnetic (EM) energy measurements
with high granularity.
A steel/scintillator-tile hadron calorimeter covers the central pseudorapidity range (\(|\eta| < 1.7\)).
The endcap and forward regions are instrumented with LAr calorimeters for both the EM and hadronic energy measurements up to \(|\eta| = 4.9\).
The MS surrounds the calorimeters and is based on three large superconducting air-core toroidal magnets with eight coils each.
The field integral of the toroids ranges between \num{2.0} and \SI{6.0}{\tesla\cdot\metre} across most of the detector.
The muon spectrometer includes a system of precision chambers for tracking and fast detectors for triggering.
A two-level trigger system is used to select events.
The first-level trigger is implemented in hardware and uses a subset of the detector information to accept events at a rate below \SI{100}{\kHz}.
This is followed by a software-based trigger that
reduces the accepted event rate to \SI{1.25}{\kHz} on average depending on the data-taking conditions.
A software suite~\cite{ATL-SOFT-PUB-2021-001} is used in data simulation, in the reconstruction and analysis of real
and simulated data, in detector operations, and in the trigger and data acquisition systems of the experiment.


%

%
\section{Data and MC simulation}
\label{sec:data_and_mc}


This measurement uses \lumi of $\sqrt{s}=\SI{13}{\TeV}$ proton--proton collision data collected by the ATLAS detector from 2015 to 2018~\cite{DAPR-2021-01}.
The selected data were collected during periods with stable LHC beam conditions and a fully functioning ATLAS detector~\cite{DAPR-2018-01}. Events were required to satisfy a single-electron or single-muon trigger~\cite{TRIG-2016-01,TRIG-2018-05}.
Single-lepton triggers with low transverse momentum (\pT) thresholds and isolation requirements were combined with triggers applying higher \pT thresholds, looser identification criteria, and no isolation requirements, resulting in an efficiency of almost 100\% for events satisfying the analysis selection criteria.
The lowest \pT threshold ranged from \SI{24}{\GeV} to \SI{26}{\GeV} for single-electron triggers and from \SI{20}{\GeV} to \SI{26}{\GeV} for single-muon triggers.
The average number of collisions per bunch crossing (pile-up) ranged from 13 to 38 during the 2015--2018 data-taking periods.

Monte Carlo (MC) simulated event samples are used to model the signal and irreducible backgrounds. The MC events were passed through a $\GEANT$-based simulation of the ATLAS detector~\cite{Agostinelli:2002hh, SOFT-2010-01} to account for detector resolution and acceptance effects. For some samples used in the evaluation of systematic uncertainties, a faster parameterised shower simulation was employed in place of the full $\GEANT$ model. Inelastic collisions were simulated using the \PYTHIA[8.186]~\cite{Sjostrand:2007gs} MC event generator with the A3 set of tuned parameters~\cite{ATL-PHYS-PUB-2016-017} and the \NNPDF[2.3lo]~\cite{Ball:2012cx} set of parton distribution functions (PDFs) and overlaid on simulated signal and background events. Correction factors are applied to account for differences between data and simulation in the amount of pile-up and in the reconstruction, particle identification, and trigger efficiencies.

Signal \wwy events including up to two light-flavour jets were generated using \SHERPA[2.2.11]~\cite{Bothmann:2019yzt} and the \NNPDF[3.0nnlo] PDF set~\cite{Ball:2014uwa}, together with the default \SHERPA tuning of parton-shower parameters.
The $\wwy$\,+\,0-jet events were simulated at next-to-leading order (NLO) in QCD, and combined with $\wwy$\,+\,1-jet and $\wwy$\,+\,2-jet event samples, which were simulated at leading order (LO) using matrix elements from the Comix~\cite{Gleisberg:2008fv} and \OPENLOOPS~\cite{Buccioni:2019sur,Cascioli:2011va,Denner:2016kdg} libraries.
The partons were matched with the \SHERPA parton shower~\cite{Schumann:2007mg} using the ME+PS@NLO prescription~\cite{Hoeche:2011fd,Hoeche:2012yf,Catani:2001cc,Hoeche:2009rj} with the set of tuned parameters developed by the \SHERPA authors.
The \wwy process was simulated in the $\ell\nu\ell\nu\gamma$ ($\ell=e,\mu,\tau$) final state, including contributions from off-shell \PW bosons and lepton pairs not originating from \PW boson decays such as $\zzy\to\ell\ell\nu\nu\gamma$.
Photons radiated from initial- or final-state charged particles were included. A minimum photon energy of \SI{7}{\GeV} and a separation of $\Delta R > 0.1$ between the photon and charged leptons were required. Processes with a $\Pqb$-quark in the initial or final state are excluded because they are dominated by top-quark contributions rather than diagrams with gauge-boson couplings.
This procedure also removes small $\Pqb\bar{\Pqb}$-induced signal contributions that are free of top quarks, but the impact on the total cross-section is less than 1\%, as evaluated with \SHERPA for $\wwy$\,+\,0-jet production at LO in QCD.

For the EFT interpretation described in \Cref{sec:EFT}, MC samples were generated with non-zero Wilson coefficients for the  \EFTOperator{M0}, \EFTOperator{M1}, \EFTOperator{M2}, \EFTOperator{M3}, \EFTOperator{M4}, \EFTOperator{M5}, \EFTOperator{M7}, \EFTOperator{T0}, \EFTOperator{T1}, \EFTOperator{T2}, \EFTOperator{T5}, \EFTOperator{T6}, and \EFTOperator{T7} operators in the Eboli model~\cite{Eboli_EFT} at LO in QCD using \MGNLO[2.9.9]~\cite{Alwall:2014hca}, interfaced with \PYTHIA[8.3], and employing the \NNPDF[3.0nnlo] PDF set. These samples include only on-shell \PW bosons and no additional partons in the matrix element (ME). The Wilson coefficients were set to $\SI{1}{\TeV^{-4}}$ in all EFT samples. In addition, a SM \wwy event sample was generated using the same \MGNLOPY set-up. The ratio of \SHERPA to \MGNLO SM predictions was used to derive $\pT(\gamma)$-dependent corrections for the EFT samples, with a correction factor of approximately \num{2.3} observed in the high-$\pT(\gamma)$ region, $\pT(\gamma) > \SI{500}{\GeV}$.

Simulated event samples containing prompt photons and leptons are used to model background processes with final states similar to the signal.
These include \ttgamma, \twy, \zy, \wzy and \zzy productions.

MC \ttgamma background events with photons from ISR were generated at NLO in QCD with \MGNLO[2.7.3]~\cite{Alwall:2014hca} using the \NNPDF[3.0nlo] PDF set~\cite{Ball:2014uwa} interfaced to \PYTHIA[8.240]~\cite{Sjostrand:2014zea}, which used the A14 set tune~\cite{ATL-PHYS-PUB-2014-021} and the \NNPDF[2.3lo] PDF set for the parton shower (PS). Top quarks were decayed at LO in QCD using \MADSPIN~\cite{Frixione:2007zp,Artoisenet:2012st}. With the same setup, a second \ttgamma sample with photons from FSR was generated as a $2 \to 2$ process with on-shell top quarks at LO precision, where the photon originates from the charged decay products of the top quarks. The normalisation of this sample is corrected by a NLO/LO inclusive $k$-factor of 1.5 obtained in Ref.~\cite{TOPQ-2021-29}.
The production of \twy was simulated at LO in QCD with \MGNLO interfaced to \PYTHIA[8.212] using the A14 tune and the \NNPDF[2.3lo] PDF set. Two samples were generated: one where the photon originates from the initial state or from radiative $tW$ production, and another where the photon arises from the decay of the top quark, the \PW boson, or their charged decay products.

The \zy MC event sample was generated as $\ell\ell\gamma$ ($\ell=e,\mu,\tau$) with \SHERPA[2.2.8]~\cite{Bothmann:2019yzt}. The \zy events with two or three light-flavour jets were simulated at LO in QCD, whereas $\zy$\,+\,0,1-jet events were simulated at NLO in QCD. Background \wzy and \zzy (referred to as \vzy) MC events were generated at NLO in QCD as $\ell\nu\ell\ell\gamma$ and $\ell\ell\ell\ell\gamma$ ($\ell=e,\mu,\tau$) final states with \SHERPA[2.2.11]. Those with one or two additional light-flavour jets were simulated at LO in QCD and combined with the $\wzy$\,+\,0-jet and $\PZ\PZ\gamma$\,+\,0-jet NLO samples. The \NNPDF[3.0nnlo] PDF set~\cite{Ball:2014uwa} was used for these samples. Events with off-shell \PW or \PZ bosons were included, as were those with a photon radiated from an initial- or final-state charged particle.

Background events in which at least one lepton or photon is not prompt or is due to misidentification, referred to as fake-lepton and fake-photon backgrounds, can also mimic the signal. These contributions are estimated using MC samples, with their normalisation and shape corrected using data-driven methods and dedicated control regions. The contributing processes include diboson ($VV$), \ttbar, \tw and \Zjets production.

The diboson samples were generated with \SHERPA[2.2.12] using the same setup as the signal, i.e.\ the \NNPDF[3.0nnlo] PDF set and the dedicated parton-shower tuning provided by the \SHERPA authors.
The \ttbar and \tw event samples were generated at NLO in QCD with \POWHEGBOX[v2]~\cite{Frixione:2007nw,Nason:2004rx,Frixione:2007vw,Alioli:2010xd} with the \NNPDF[3.0nlo] PDF set and the \hdamp\ parameter\footnote{The \hdamp\ parameter is a resummation damping factor and one of the parameters that controls the matching of the \POWHEGBOX ME to the PS and thus effectively regulates the high-\pT\ radiation against which the \ttbar\ system recoils.} set to 1.5\,\mtop~\cite{ATL-PHYS-PUB-2016-020}. Both samples were interfaced to \PYTHIA[8.230]~\cite{Sjostrand:2014zea}, which used the A14 tune and the \NNPDF[2.3lo] PDF set.
The \Zjets MC sample was generated with \SHERPA[2.2.1]~\cite{Bothmann:2019yzt}, using NLO QCD matrix elements for up to two partons and LO QCD matrix elements for three or four partons, together with the \NNPDF[3.0nnlo] PDF set.


%

%
\section{Object reconstruction and event selection}
%


A primary vertex~\cite{ATL-PHYS-PUB-2015-026} is constructed by requiring at least two tracks consistent with a $pp$ collision at the interaction point. The primary vertex with the highest sum of the squared transverse momenta of all associated tracks with $\pT>\SI{500}{\MeV}$ is selected.

Electron candidates are reconstructed from energy clusters in the EM calorimeter and matched to reconstructed tracks in the ID~\cite{EGAM-2018-01,PERF-2015-08}. Electrons are also required to meet the \texttt{TightLH} identification and \texttt{PLVTight} isolation criteria, and to satisfy $\pT > \SI{20}{\GeV}$ and $\abseta < 2.47$, excluding the transition region $1.37 < \abseta < 1.52$ between the calorimeter barrel and endcaps. Electron tracks must also be consistent with originating from the primary vertex by satisfying $|d_0|/\sigma_{d_0}<5$ and $|z_0 \cdot \sin\theta|<\SI{0.5}{\mm}$. Here, $d_0$ denotes the transverse impact parameter with respect to the beamline, \( \sigma_{d_0} \) is its associated uncertainty, $z_0$ is the longitudinal impact parameter with respect to the primary vertex, and \( \theta \) is the polar angle of the track.

Photon candidates are reconstructed from energy clusters in the EM calorimeter~\cite{EGAM-2018-01}. Photon energy clusters must be within $\abseta< 2.37$, excluding the transition region $1.37 < \abseta < 1.52$ between the barrel and endcaps. Photon candidates are required to meet the \texttt{Tight} identification and \texttt{FixedCutTight} isolation criteria and have $\pT > \SI{20}{\GeV}$~\cite{EGAM-2018-01}.

Muon candidates are reconstructed by combining a track reconstructed in the ID with one reconstructed in the MS~\cite{MUON-2018-03} and are required to have $\pT>\SI{20}{\GeV}$ and $\abseta<2.5$. Several quality requirements are imposed to reject non-prompt muon candidates, which originate primarily from pion and kaon decays. Muons are required to meet \texttt{Medium} identification and \texttt{Tight} isolation criteria~\cite{MUON-2018-03}. Muon tracks are also required to be consistent with originating from the primary vertex by satisfying $|d_0|/\sigma_{d_0} < 3$ and $|z_0\cdot\sin\theta| < \SI{0.5}{\mm}$.

Jet-related variables play an important role in this analysis. A veto on jets originating from \Pqb-quarks (\Pqb-jets) is primarily used to suppress backgrounds from top-quark processes, and additional jet-related observables are used in the boosted decision tree to enhance signal--background separation.
Jets are reconstructed from particle-flow objects~\cite{PERF-2015-09}, using the anti-$k_t$ algorithm~\cite{Cacciari:2008gp,Fastjet} with a radius parameter of $R=0.4$. Jets are required to have $\pT > \SI{20}{\GeV}$ and $\abseta < 2.5$. Jets with $\pT < \SI{60}{\GeV}$ and $\abseta < 2.5$ must fulfil the requirement of a likelihood tagger called the Jet Vertex Tagger (\texttt{JVT}) to reduce the impact of pile-up~\cite{ATLAS-CONF-2014-018}. A \texttt{JVT} requirement that has 92\% efficiency while rejecting 98\% of jets from pile-up and noise is imposed.
Identification of \Pqb-jets is based on reconstructing secondary vertices formed by tracks associated with jets and combining their spatial-parameter values with lifetime-related information. The \texttt{DL1r} \Pqb-tagging algorithm~\cite{FTAG-2018-01,FTAG-2019-02,FTAG-2019-07} is used at a working point with an \SI{85}{\%} \Pqb-jet-tagging efficiency, independent of \Pqb-jet \pT and calibrated using simulated \ttbar events. The rejection rates for light-flavour jets and \Pqc-jets are \pT-dependent, and on average are 40 and 3, respectively.

The missing transverse momentum, \MET, provides a measure of the transverse momentum imbalance due to escaping neutrinos. The \MET is calculated as the magnitude of the negative vector sum of the transverse momenta of all reconstructed electrons, muons, photons, and jets in the event, plus a term to account for soft hadronic activity, built from tracks associated with the hard-scattering vertex but not matched to any of the reconstructed objects~\cite{PERF-2016-07}.

Object reconstruction is followed by an overlap removal procedure applied to electrons, photons, muons and jets to remove possible duplication due to individual physical objects being reconstructed as two different entities in the detector~\cite{TOPQ-2019-19}.

The \wwy process yields several different final states depending on the decay modes of the \PW bosons. In this measurement, the targeted final state contains an electron and a muon of opposite electric charge, accompanied by missing transverse momentum and exactly one photon. This selection defines the \emuy signal region (SR). The \emuy final state, together with a veto on events containing \Pqb-jets, is chosen in order to suppress background contributions, particularly from \tty, \zy, and \twy production. Throughout this analysis, $\ell$ refers exclusively to electrons or muons.

Events must have exactly one muon and one electron satisfying the requirements described above.
At least one lepton must have $\pT>\SI{27}{\GeV}$ to satisfy the trigger threshold, and be matched to the lepton that was accepted by the trigger. Exactly one photon is required, meeting the identification and isolation requirements specified previously. In addition, events are required to have $\MET>\SI{20}{\GeV}$. A $\PZ$-boson veto is applied mainly to reduce contributions from $\PW\PZ\rightarrow \mu^{\pm}\nu \ee$, where one electron from the $\PZ\rightarrow \ee$ decay is reconstructed as a prompt photon. This is achieved by requiring $|\mey - m_{\PZ}| > \SI{5}{\GeV}$. Additionally, events containing \Pqb-jets are vetoed to reduce \tty and \twy background contributions. These signal event requirements are summarised in~\Cref{tab:region_defs}.


%

\section{Background estimation}
\label{sec:background}


The background yields are estimated using a combination of MC simulations and data-driven methods, and are constrained using statistically independent control regions (CRs). \Cref{tab:region_defs} summarises the event selection criteria and defines the CRs used to estimate the normalisation of the major backgrounds.

\begin{table}[ht!]
\centering
\caption{Signal and control regions used in the analysis. The \PZ-boson mass $m_{\PZ}$ is taken to be $\SI{90}{\GeV}$. Selections shown in boldface correspond to differences with respect to the SR, chosen both to enhance the targeted background and to ensure orthogonality between regions.}
\label{tab:region_defs}
\footnotesize
\begin{tabular}{ccccc}
\toprule
\multicolumn{5}{c}{Common selections}\\
\midrule
\multicolumn{5}{c}{$\text{Leading lepton } \pt  > \SI{27}{\GeV}$} \\
\multicolumn{5}{c}{$\text{Subleading lepton } \pt  > \SI{20}{\GeV}$} \\
\multicolumn{5}{c}{Exactly 1 photon with \pT > \SI{20}{\GeV}} \\
\multicolumn{5}{c}{$\MET  > \SI{20}{\GeV}$} \\
\midrule
SR \emuy & CR \tty & CR1 \zy & CR2 \zy & CR \efake \\
\midrule
\DFOS & \DFOS & \textbf{\SFOS} & \textbf{\SFOS}  & \DFOS\\
$|\mey - m_{\PZ}| > \SI{5}{\GeV}$ & $|\mey - m_{\PZ}| > \SI{5}{\GeV}$ &
\textbf{--} & \textbf{--}  & \textbf{$|\mey - m_{\PZ}| < \SI{5}{\GeV}$}  \\
-- & -- & \textbf{$m(\ell\ell\gamma)< \SI{100}{\GeV}$} & \textbf{$m(\ell\ell\gamma)> \SI{100}{\GeV}$} & -- \\
0 \Pqb-jets & \textbf{1 \Pqb-jet} & 0 \Pqb-jets & 0 \Pqb-jets & 0 \Pqb-jets\\
\bottomrule
\end{tabular}
\end{table}

The modelling of prompt backgrounds, such as those due to the \tty and \zy processes, is based on MC simulations, with their normalisations determined by a maximum-likelihood fit using dedicated \tty and \zy CRs. The \tty CR matches the kinematic properties of the signal region, but requires exactly one \Pqb-jet.
Two \zy CRs are constructed, each selecting $\ell\ell\gamma$ events with a \SFOS lepton pair. The \zy CR1 requires $\mlly < \SI{100}{\GeV}$, whereas \zy CR2 requires $\mlly > \SI{100}{\GeV}$, resulting in an approximately 30\% contribution from jets or neutral hadrons misidentified as photons (\jfake), making it well suited to constrain this background. An additional control region (the \efake CR) is defined by requiring $|m_{e\gamma} - m_{\PZ}| < \SI{5}{\GeV}$, in order to constrain backgrounds from electrons misidentified as photons (\efake). This region is dominated by $\PW\PZ$ events, where an electron from the \PZ-boson decay is reconstructed as a photon.

These \jfake background events stem from processes without prompt photons, typically \ttbar, \Zjets, or diboson ($VV$), where jets occasionally satisfy the photon identification and isolation criteria.
Samples of MC events from these processes are used to estimate the \jfake misidentification background. An overlap-removal procedure ensures orthogonality between prompt- and non-prompt-photon MC samples~\cite{TOPQ-2018-44}. Events with photons from initial- or final-state radiation (ISR/FSR) are included in prompt-photon samples (e.g.\ \tty), while jet-induced photon-like objects remain in non-prompt-photon samples (e.g.\ \ttbar). Discrepancies between MC simulation and data are minimised by applying two-dimensional scale factors (SFs) derived as functions of the scalar sum of jet transverse momenta, $\Sigma \pT(\text{jet})$, and the transverse mass of the dilepton--\MET system, $m_\mathrm{T}(\ell\ell,\MET)$. The use of these variables, sensitive to event kinematics and the presence of jets and leptons, effectively addresses modelling mismatches.

The correction SFs are calculated in a statistically independent region enriched in \jfake events, where SR selection requirements other than the photon isolation criterion are applied. The photon candidate fails the \texttt{FixedCutTight} isolation requirement~\cite{EGAM-2018-01}, but passes a looser selection, in which the sum of track \pT within a cone of $\Delta R=0.2$ around the photon is less than 5\% of $\pT(\gamma)$, and the calorimeter-based isolation requirement is reversed. This region is kinematically similar to the SR, allowing the scale factors to be applied reliably. The SFs are computed by normalising the \jfake MC event yields to data after subtracting contributions from prompt backgrounds. The resulting scale factors typically range between 0.8 and 1.5. The associated uncertainties, arising from limited numbers of data and MC events, range from $\pm0.2$ to $\pm0.7$ and are treated as additional systematic uncertainties.

The \jfake background estimate is cross-checked using a two-dimensional sideband technique~\cite{STDM-2018-33}. Four non-overlapping regions are defined by using two uncorrelated variables:\ photon identification and photon isolation. The SR requires photons to satisfy both the \texttt{Tight} identification and the \texttt{FixedCutTight} isolation criteria. Three CRs are constructed by inverting the identification and/or isolation criteria to enhance the contribution from misidentified jets. This is done by selecting events with a photon candidate that passes the \texttt{Loose} photon identification requirement but fails at least two of the EM shower-shape requirements used in \texttt{Tight} identification; these photons are denoted by $T'$. For isolation ($I$), the photon candidate must pass the track-isolation requirement but fail the calorimeter-isolation requirement ($I'$). The \jfake yield in the SR is then estimated as $TI = T'I \cdot TI' / T'I'$. As an additional cross-check, the procedure is repeated with photon identification paired with the requirement of same-sign lepton events instead of the \DFOS selection.
The two \jfake estimates, obtained from the sideband and scale-factor methods, differ by 30\%, which is assigned as an additional systematic uncertainty on the \jfake background in the SR.

Backgrounds such as diboson events with electron-to-photon (\efake) misidentification are significantly suppressed by rejecting events with \mey within the \PZ-boson mass window. The remaining \efake \enquote{fake rate} is estimated using a data-driven reweighting method, as described in Refs.~\cite{TOPQ-2018-24, STDM-2018-33}.
The rate estimate is derived from $Z\rightarrow e^{+}e^{-}$ events reconstructed as $e\gamma$ final states, using two dedicated regions.
An $e\gamma$ region is defined by requiring one electron and one photon with an invariant mass within \SI{35}{\GeV} of the \PZ-boson mass, and the photon \pT to be less than the electron \pT (to make the photon more likely to be fake). An $e^{+}e^{-}$ region is defined by selecting events with two oppositely charged electrons with an invariant mass within \SI{35}{\GeV} of the \PZ-boson mass. In both regions, events containing $\Pqb$-jets are removed.

The invariant mass distributions in the $e\gamma$ and $e^{+}e^{-}$ control regions are modelled using a single-sided Crystal Ball function~\cite{Das:2016stf} for the \PZ-boson resonance and a fifth-order Bernstein polynomial for non-resonant background processes.
The yields $N_{e\gamma}$ and $N_{ee}$ are binned in \pT and $|\eta|$ of the fake photon, and fake-factors are then defined as the ratio $N_{e\gamma}/N_{ee}$.
The correction scale factors, typically between 0.71 and 1.26, are derived as the MC-to-data ratio of the fake-factors, and are used to reweight MC samples that include \efake misidentification events.
A systematic uncertainty is assigned to account for the limited number of data and MC events used in the evaluation of the scale factors, ranging from $\pm0.01$ to $\pm0.21$. An additional background uncertainty is included to account for the choice of parameterisation.

Finally, contributions from fake leptons are found to be negligible in the regions considered. Photon contributions from pile-up are evaluated following the method described in Ref.~\cite{STDM-2020-14} and are found to be well modelled in the MC simulation, with no additional treatment required.

The predicted yield of \wwy events in the SR after the likelihood fit (post-fit) is $250 \pm 40$, constituting 26\% of the expected total number of events. The quoted uncertainty includes both statistical and systematic components from the likelihood fit to data, as described in \Cref{sec:result}.
Most of the background in the \emuy SR comes from sources with prompt photons, including \tty (45\%), \zy (13\%) and \vzy (3\%).
The remaining background contributions are due to non-prompt or misidentified photons, and include events where an electron is misidentified as a photon (\efake, 3.5\%) and events where a jet is misidentified as a photon (\jfake, 9.5\%).


%
\section{Multivariate analysis}
%


To improve the separation between signal and background events, the \xgb~\cite{Chen:2016btl} package is used to train a boosted decision tree (BDT), which constructs a discriminant that takes correlations among the input variables into account, resulting in a single continuous output variable in the interval $[0,1]$.
The BDT is implemented with hyperparameters optimised for performance and stability, using 120 trees with a maximum depth of four.

The BDT is trained using MC events from the SR, where all signal and background processes are included in the training and weighted according to their expected event yields. The $\efake$ and $\jfake$ backgrounds are scaled using the MC correction scale factors described in~\Cref{sec:background}.
A 5-fold cross-validation strategy \cite{Stone1974,James2013} is used to improve the statistical stability of the trained BDT by training five independent models. The resulting BDT output distributions from the five held-out sets are added to form the final BDT distribution used in the statistical fit.

Only variables that provide good signal–background separation and are well modelled are used in the BDT training.
The ten input variables with the highest feature importance are selected, and the full list is shown in~\Cref{tab:training_variables}.
The feature importance is calculated using the Gain method~\cite{Chen:2016btl}.

\begin{table}[h!]
\begin{center}
\caption{
Input variables used to train the BDT in the \emuy SR, ordered by feature importance.
For jet-related observables, values are set to $-99$ if the required number of jets is not reconstructed: $\Delta R(jj)$ and $\pT(j_{2})$ for fewer than two jets, and $\pT(j_{1})$ for zero jets.}
\label{tab:training_variables}
\begin{tabular}{lcl}
\toprule
\multirow{2}{*}{Variable} & Feature & Definition\\
& importance \\
\midrule

$m_\text{T}(\ell_{1}, \MET)$ & 13\% & Transverse mass of the leading lepton and \MET. \\
$\Delta R(jj)$ & 11\% & $\Delta R$ between the leading and subleading jets. \\
$m(\ell\ell)$ & 11\% & Invariant mass of the two-lepton system in the event. \\
$\pT(j_{2})$ & 11\% & \pT of the subleading jet. \\
$\Sigma |\pT(j)|$ & 10\% & Scalar sum of the \pT of all jets in the event. \\
$m_\text{T}(\ell_{2}, \MET)$ & 10\% & Transverse mass of the subleading lepton and \MET. \\
$\pT(j_{1})$ & 10\% & \pT of the leading jet. \\
$N(\text{jets})$ & ~~9\% & Number of jets in the event. \\
$\Sigma |\pT(\ell)|$ & ~~9\% & Scalar sum of the \pT of the two leptons in the event. \\
$m(\ell_2\gamma)$ & ~~6\% & Invariant mass of the subleading lepton and photon. \\

\bottomrule
\end{tabular}
\end{center}
\end{table}

\begin{figure}[htbp]
\centering
\begin{subfigure}[b]{0.35\textwidth}%
\includegraphics[width=\textwidth]{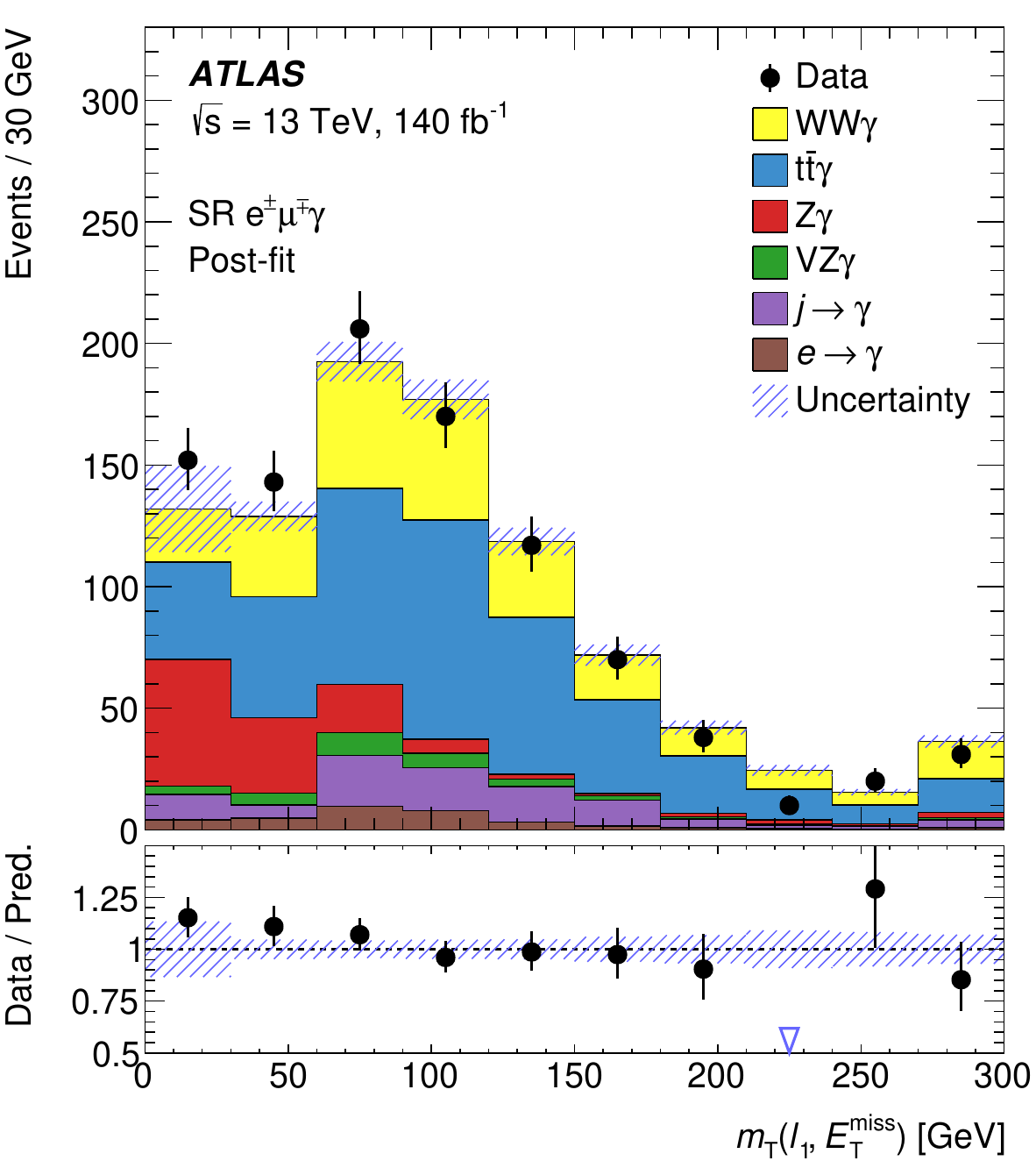}
\caption{}
\end{subfigure}%
\begin{subfigure}[b]{0.35\textwidth}%
\includegraphics[width=\textwidth]{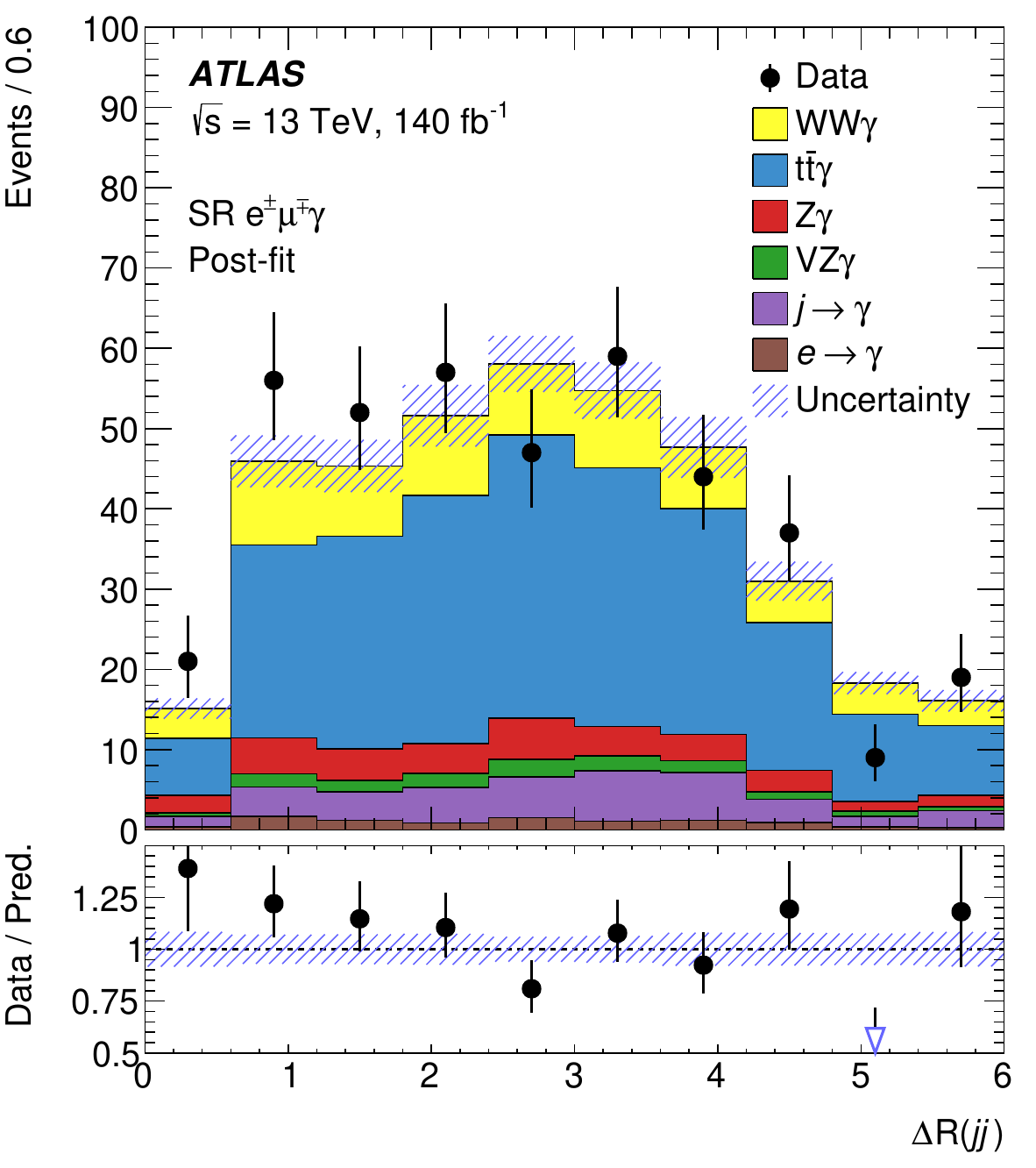}
\caption{}
\end{subfigure} \\
\begin{subfigure}[b]{0.35\textwidth}%
\includegraphics[width=\textwidth]{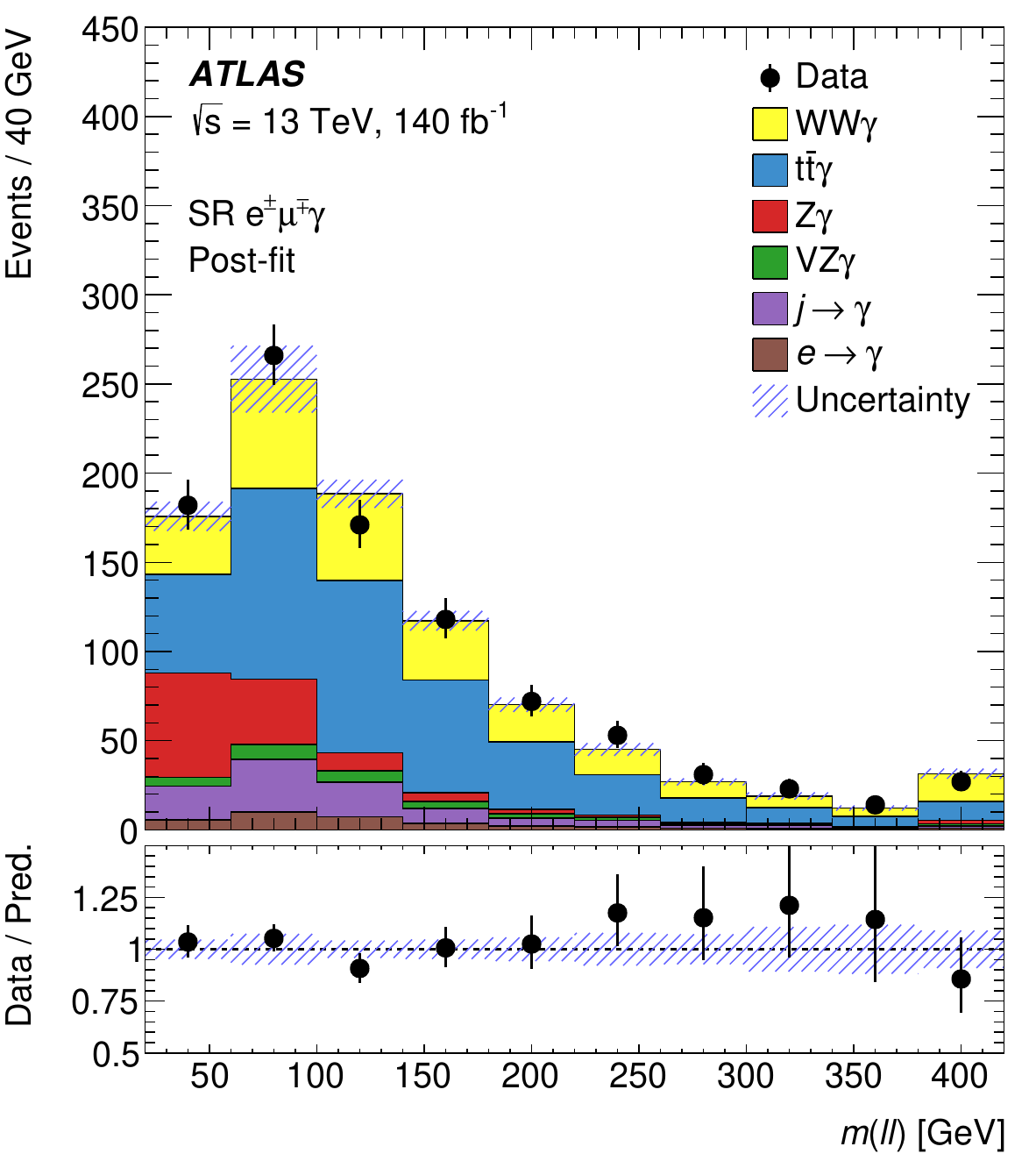}
\caption{}
\end{subfigure}
\begin{subfigure}[b]{0.35\textwidth}%
\includegraphics[width=\textwidth]{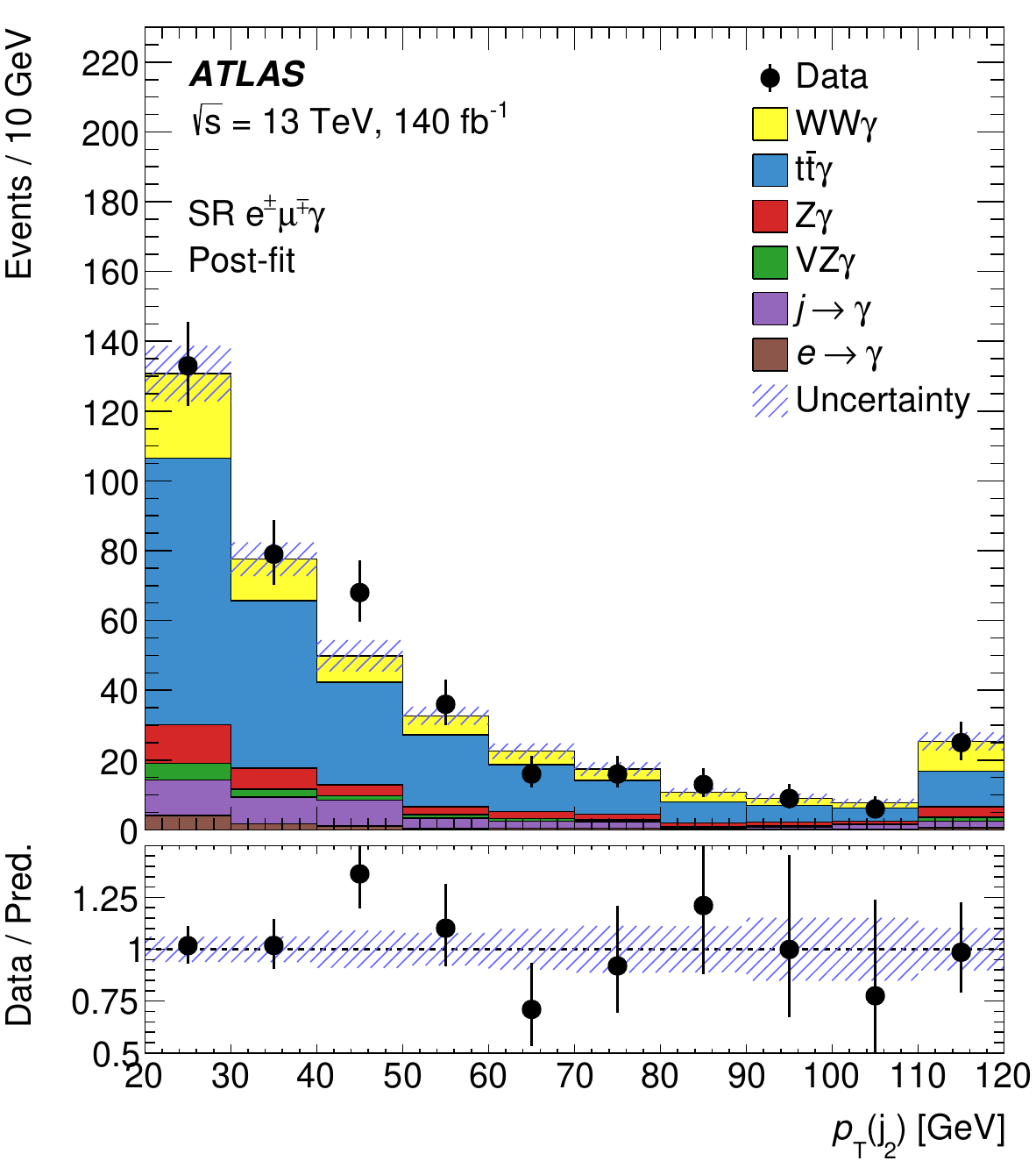}
\caption{}
\end{subfigure}%

\caption{
Comparison between data and prediction (\enquote{Pred.}) for the four most discriminating BDT input variables after the fit to data (\enquote{Post-Fit}) under the signal-plus-background hypothesis, shown in the SR.
The variables displayed are (a) $m_\text{T}(\ell_{1}, \MET)$, (b) $\Delta R(jj)$, (c) $m(\ell\ell)$, and (d) $\pT(j_{2})$. The uncertainty band includes both the statistical and systematic uncertainties as obtained from the fit. The rightmost bin includes overflow events. The lower panels show the ratio of data to prediction. Open markers indicate data points lying outside the vertical range of the plot.}

\label{fig:emuysr_plots}
\end{figure}

The distributions of the four variables contributing the most to the BDT, shown after the likelihood fit described in~\Cref{sec:result}, are presented in \Cref{fig:emuysr_plots}. The jet-related variables provide good separation between the \wwy and \tty processes. In contrast, the remaining BDT input variables provide discrimination power between the signal and the \zy and \vzy backgrounds.

%

%
\section{Systematic uncertainties}
\label{sec:systematic_uncert}


Systematic uncertainties related to the efficiency of reconstructing the \wwy signal and major backgrounds, and the normalisation of the \vzy background contribution, are taken into account. Experimental and theoretical uncertainties affecting the kinematic distributions used in the analysis are also included, with correlations across the analysis regions applied.

\paragraph{Reconstruction efficiency and calibration uncertainties:}

Systematic uncertainties affecting the reconstruction efficiency, energy calibration, and identification of electrons, photons, muons and jets are propagated through the analysis. Each source contributes only at the few-percent level to the total uncertainty in the measured signal cross-section.

Differences between electron (muon) trigger, reconstruction and selection efficiencies in data and those in MC simulation are minimised by applying scale factors derived from dedicated $\PZ \rightarrow \ee$ ($\PZ \rightarrow \mumu$) enriched control samples using a tag-and-probe method~\cite{EGAM-2018-01,MUON-2018-03}. Similarly, corrections and associated uncertainties for photon reconstruction and identification efficiencies are derived as described in Ref.~\cite{EGAM-2018-01}.

The jet energy scale (JES) was derived using test-beam data, LHC collision data, and simulation information. The JES calibration~\cite{JETM-2018-05} includes corrections that account for detector problems, jet fragmentation, the impact of pile-up, and differences between data and MC simulation.
The impact of the uncertainty in the jet energy resolution (JER)~\cite{JETM-2018-05} is evaluated by smearing the jet energy in the MC and data samples accordingly.
An additional uncertainty arises from the correction applied to simulated events associated with the \texttt{JVT} selection~\cite{PERF-2014-03}.

The \btag efficiencies and mistagging rates are measured in data as described
in Refs.~\cite{FTAG-2018-01,FTAG-2020-08,FTAG-2019-02},
with the systematic uncertainties associated with the \btag efficiency and the mistagging rates estimated separately. The impact of the uncertainties in the \btag calibration is evaluated separately for \Pqb-jets, \Pqc-jets and light-flavour jets in the MC samples.

\paragraph{\MET:} The uncertainty in \MET due to a possible miscalibration of the soft-track component of the \MET is derived from data--MC comparisons of the \pT balance between the hard and soft \MET components~\cite{PERF-2016-07}. The uncertainties associated with the leptons, jets and photons are propagated from their energy/momentum scale and resolution uncertainties, and are grouped with other uncertainties associated with those objects.

\paragraph{Luminosity:} The uncertainty in the combined 2015--2018 integrated luminosity is 0.83\%~\cite{DAPR-2021-01}, obtained using the LUCID-2 detector~\cite{LUCID2} for the primary luminosity measurements.

\paragraph{Uncertainty in \pileup modelling:} The uncertainty in \pileup modelling is accounted for by varying the reweighting of the MC samples to the data \pileup conditions by the uncertainty in the average number of interactions per bunch crossing.

\paragraph{Signal and background modelling:}

The systematic uncertainties in the \wwy and \zy modelling are estimated using the internal \SHERPA[2.2.11]  weights. These correspond to independently varying the renormalisation scale (\muR) up and down by a factor of 2 while keeping \muF fixed, and vice versa for the factorisation scale (\muF).
The PDF uncertainties are estimated using the nominal and variation weights provided for the \PDFforLHC PDF set~\cite{Butterworth:2015oua}.
The resulting uncertainty is applied to the nominal signal sample.
The uncertainty from the strong coupling constant, \alphas, is evaluated for the \wwy process, using the \NNPDF[3.0] PDF set with varied \alphas values.
The variations in \muR, \muF, the PDFs, and \alphas are applied simultaneously to both the matrix-element calculation and the parton shower.

The systematic uncertainties associated with the renormalisation and factorisation scales of the \tty and \twy processes are estimated by raising and lowering \muR and \muF by a factor of 2 in \MGNLO.
The PDF uncertainties are evaluated with the \NNPDF[2.3lo] PDF set~\cite{Ball:2012cx} following the \PDFforLHC prescription~\cite{Butterworth:2015oua}.
The systematic uncertainty associated with the parton shower and hadronisation modelling is estimated by comparing samples generated with \MGNLO[2.3.3] and interfaced with either \PYTHIA[8.212] or \Herwig[7].
For the \tty process, an additional systematic uncertainty due to initial- and final-state radiation modelling and \alphas is estimated by varying the \enquote{Var3c} parameter~\cite{ATL-PHYS-PUB-2014-021} in the A14 tune of \PYTHIA[8].

\paragraph{Data-driven background:}

For the \jfake background, a systematic uncertainty is included to account for the limited numbers of data and MC events in the regions used to derive the estimate. This uncertainty is applied separately to each MC process contributing to the estimate, i.e.\ \ttbar, \Zjets, and $VV\gamma$ as detailed in \Cref{sec:background}, and contributes approximately 2.5\% to the measurement uncertainty.

An additional uncertainty associated with the $\jfake$ misidentification is applied separately to processes involving light-flavour jets, such as \Zjets\ and $VV$, and to those involving heavy-flavour jets, such as \ttbar and $tW$. The total contribution from \jfake misidentification to the overall uncertainty is approximately 4.5\%.

A systematic uncertainty is assigned to the \efake background to account for the limited numbers of data and MC events in the $e\gamma$ and $e^{\pm}e^{\mp}$ regions.  An additional uncertainty, estimated by using the alternative non-resonant background model described in \Cref{sec:background}, affects the measurement uncertainty by less than 1\%. In total, the data-driven background estimate contributes a 5.4\% uncertainty to the measurement.

\paragraph{Background rate uncertainty:} For the \vzy process, a 20\% normalisation uncertainty is applied, motivated by the $WZ\gamma$ measurement~\cite{STDM-2019-17}. This accounts for uncertainties from the PDFs, renormalisation and factorisation scale variations, and other possible missing higher-order QCD corrections.


%
%
\section{Results}
\label{sec:result}


To extract the \wwy signal strength, a simultaneous binned maximum-likelihood fit is performed on the BDT output distributions in the \wwy SR and the \tty, \zy and \efake CRs. The fit is implemented using the \texttt{TRExFitter} framework~\cite{Aly2025TRExFitter}, which builds upon \texttt{HistFactory}~\cite{ROOSTATS} and \texttt{RooFit}~\cite{RooFit}.

The BDT trained in the SR is evaluated on the events selected in the corresponding CRs. Five unconstrained fit parameters corresponding to the \wwy, \tty, \zy, \jfake, and \efake normalisations are included.
Nuisance parameters are included in the fit to account for each systematic uncertainty described in \Cref{sec:systematic_uncert}.

\begin{figure}[ht]
\centering
\begin{subfigure}[b]{0.32\textwidth}
\includegraphics[width=\textwidth]{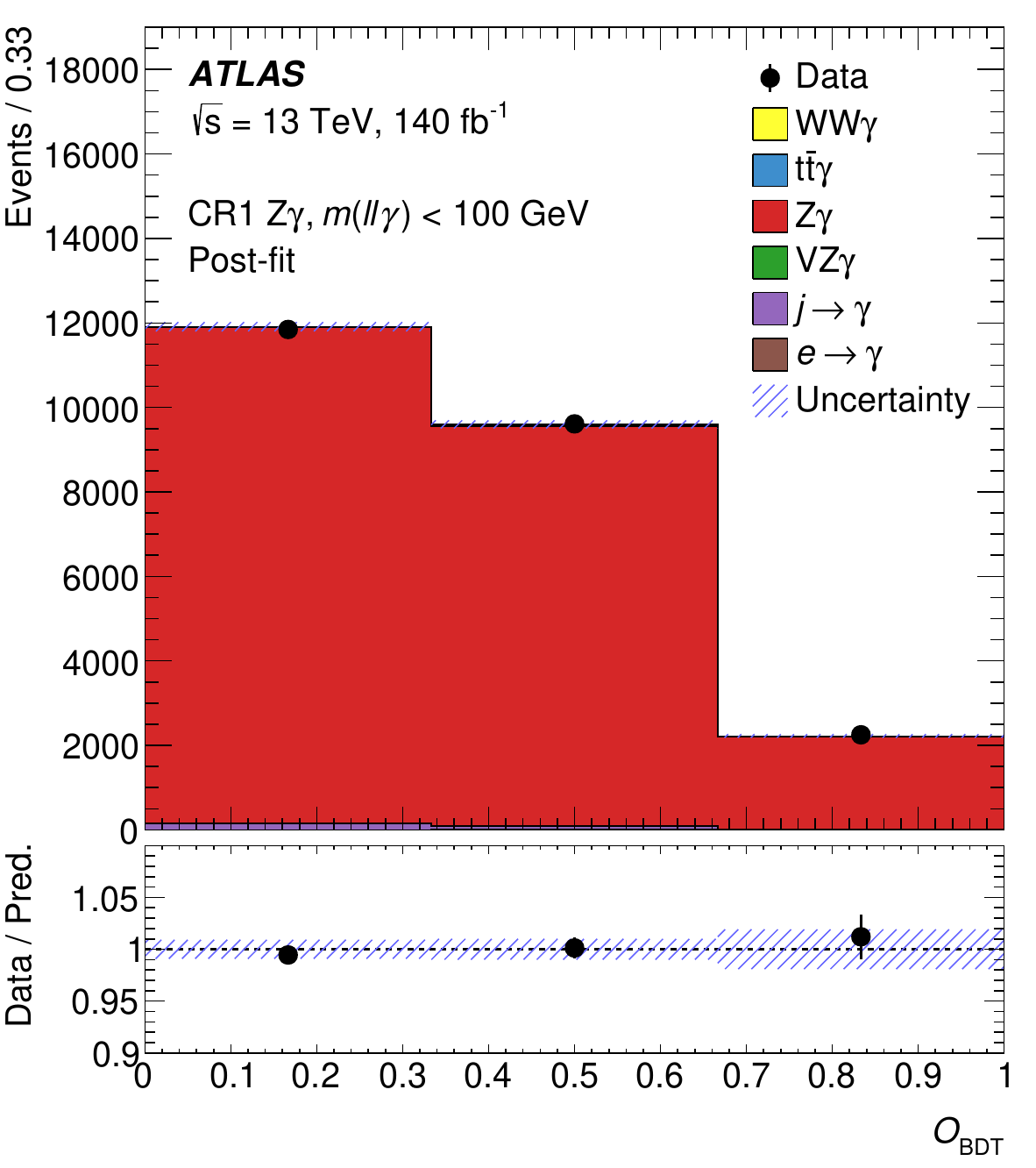}
\caption{}
\end{subfigure}
\begin{subfigure}[b]{0.32\textwidth}
\includegraphics[width=\textwidth]{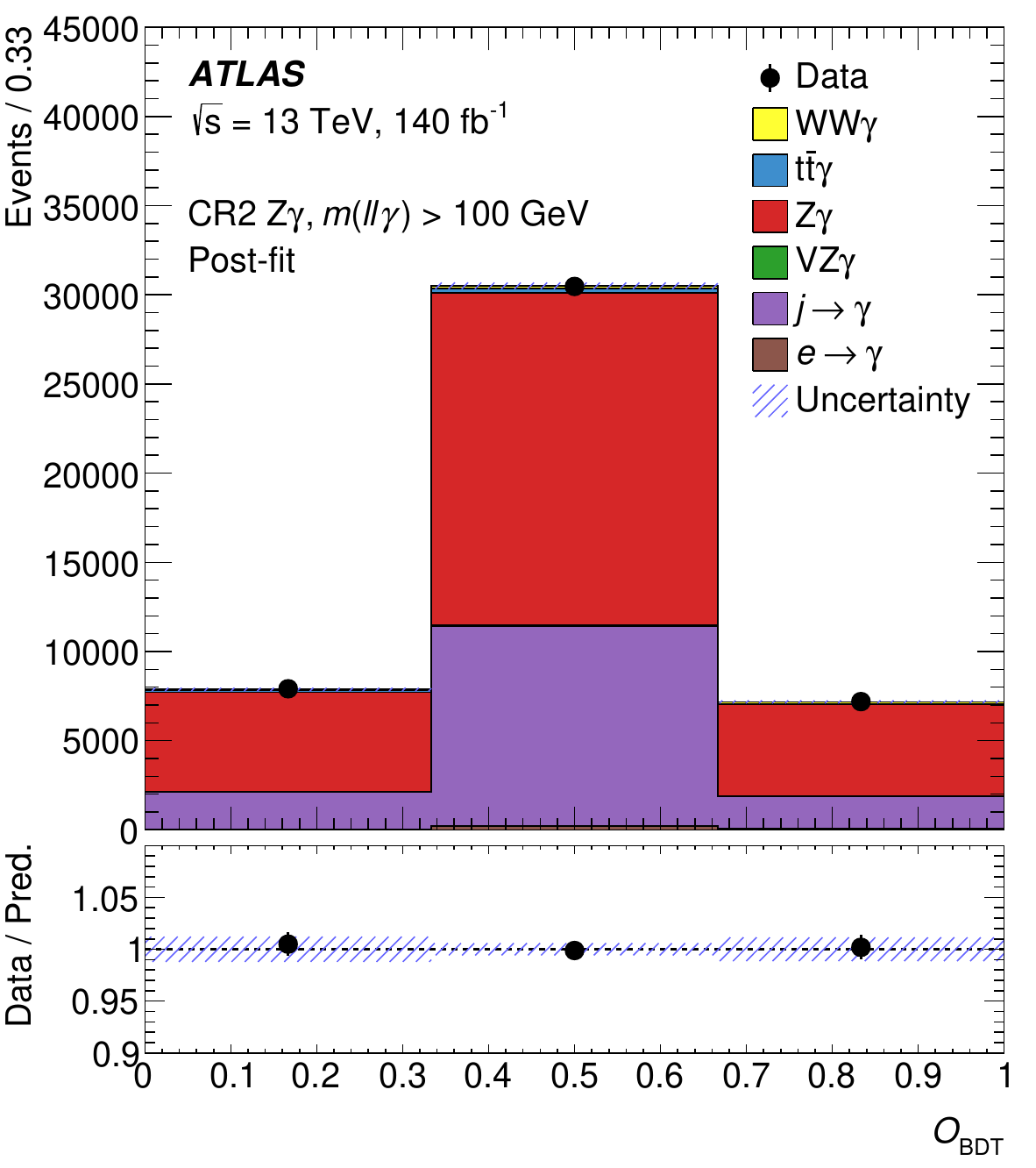}
\caption{}
\end{subfigure}
\begin{subfigure}[b]{0.32\textwidth}
\includegraphics[width=\textwidth]{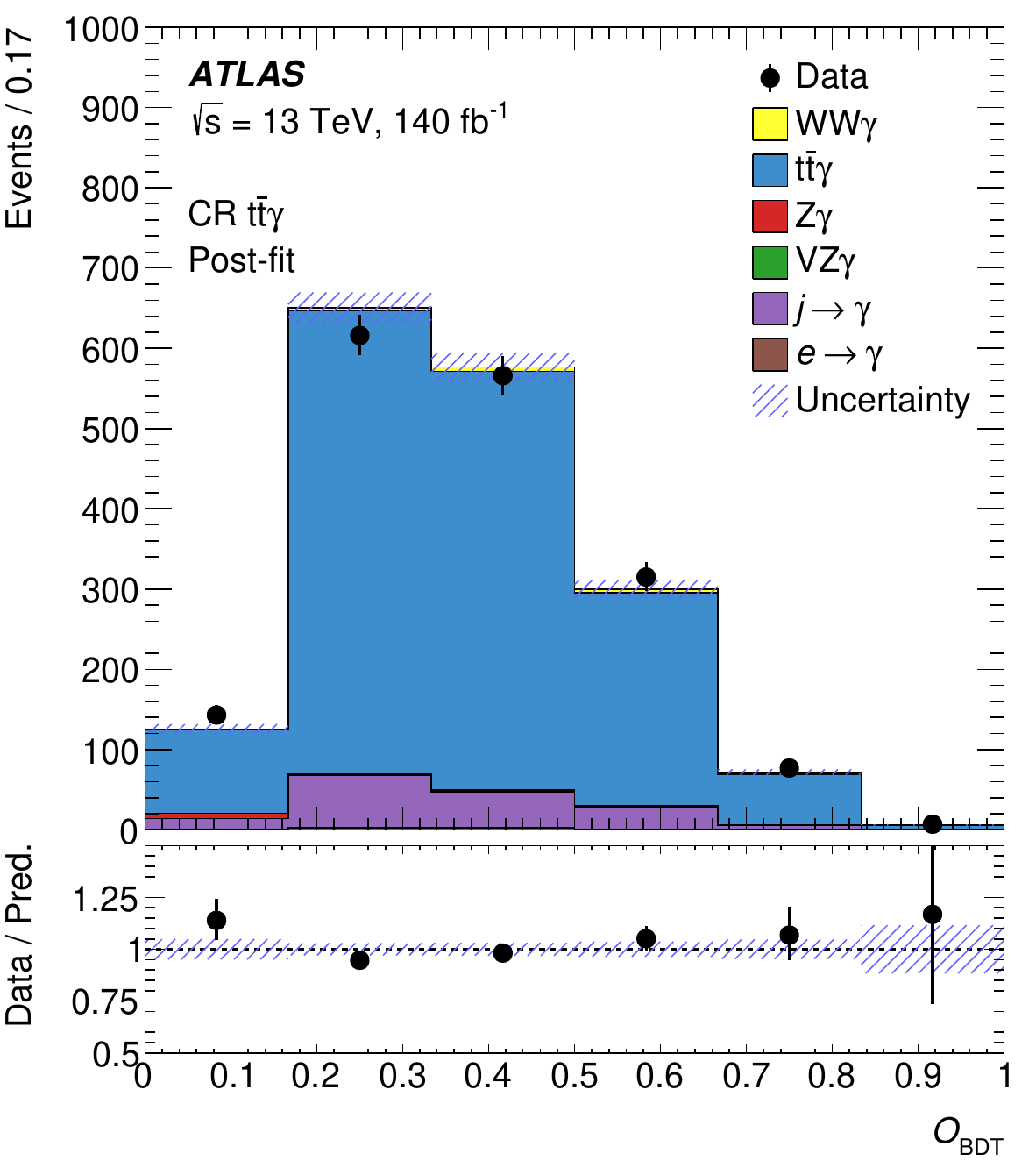}
\caption{}
\end{subfigure}\\
\begin{subfigure}[b]{0.32\textwidth}
\includegraphics[width=\textwidth]{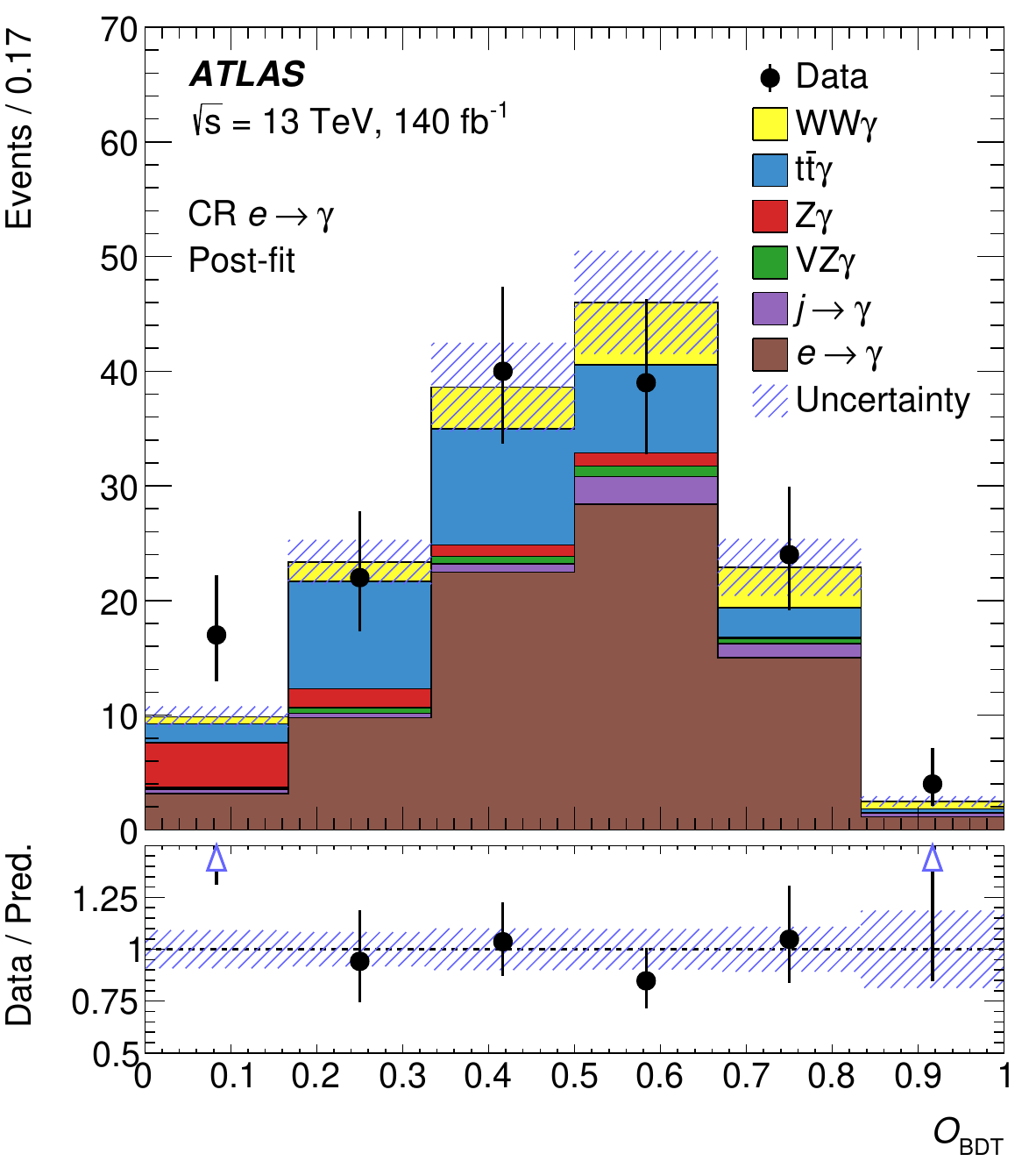}
\caption{}
\end{subfigure}
\begin{subfigure}[b]{0.32\textwidth}
\includegraphics[width=\textwidth]{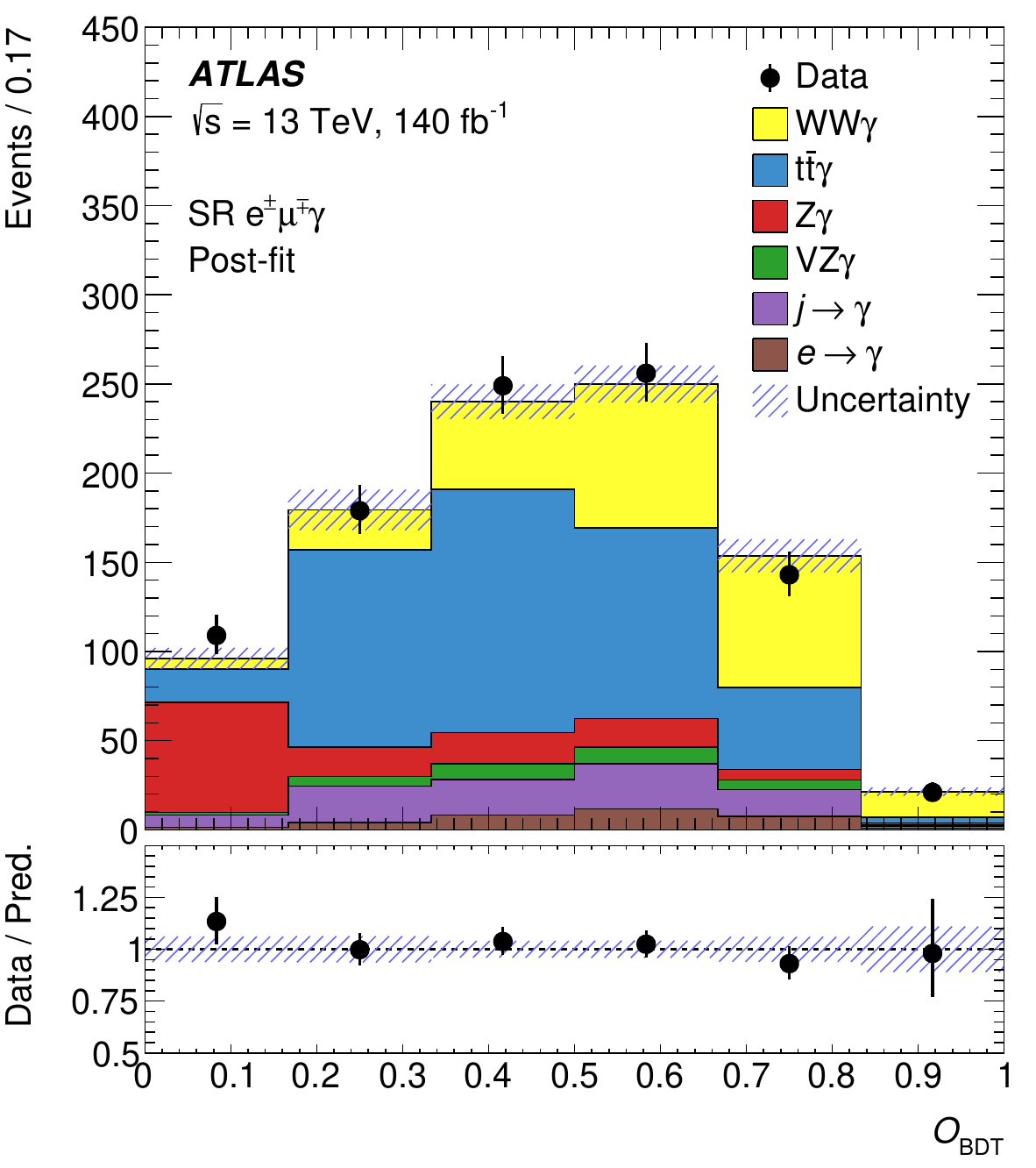}
\caption{}
\end{subfigure}

\caption{
Comparison between data and the post-fit prediction (\enquote{Pred.}), under the signal-plus-background hypothesis, from the distributions of the boosted decision tree output (\OBDT) in the (a) \zy CR1, (b) \zy CR2, (c) \tty CR, (d) \efake CR, and (e) \emuy SR. The uncertainty band includes both the statistical and systematic uncertainties as obtained from the fit.
The lower panels show the ratio of data to prediction. The open markers indicate data points lying outside the vertical range of the lower panel in (d).}
\label{fig:postfit_plots_bdt}
\end{figure}

\Cref{fig:postfit_plots_bdt} compares the BDT output distribution for the data with those for the post-fit signal and backgrounds. The numbers of fitted signal and background events are compared with the data in \Cref{tab:yields}.
The normalisation of the \wwy signal is $1.03^{\,+0.22}_{-0.21}$, in good agreement with the SM prediction, and includes the uncertainty from the predicted fiducial cross-section. The measured normalisation parameters for the backgrounds are also consistent with the SM predictions within uncertainties.

\begin{table}[htp!]

\centering
\caption{
Predicted and observed event yields in the signal region (SR) and control regions (CR). The signal and background predictions are shown after the fit to data. The uncertainties include the statistical and systematic uncertainties of the yields, and also the correlations among nuisance parameters, as well as correlations across signal and background processes because of shared systematic effects.
}
\label{tab:yields}
\sisetup{round-mode=none, scientific-notation=fixed, fixed-exponent=0, uncertainty-mode=separate, table-align-uncertainty=true}
\begin{tabular}{l
S[table-format=3.0]@{$\,\pm\,$}S[table-format=2.0, table-number-alignment=left]
S[table-format=4.1]@{$\,\pm\,$}S[table-format=2.1]
S[table-format=5.1]@{$\,\pm\,$}S[table-format=3.0]
S[table-format=5.0]@{$\,\pm\,$}S[table-format=3.0]
S[table-format=3.1]@{$\,\pm\,$}S[table-format=2.1]
}
\toprule
& \multicolumn{2}{c}{SR \emuy} & \multicolumn{2}{c}{CR $t\bar{t}\gamma$} & \multicolumn{2}{c}{CR1 \zy} & \multicolumn{2}{c}{CR2 \zy} & \multicolumn{2}{c}{ CR \efake} \\
\midrule
\wwy          & 250   & 40    & 17.7  & 3.4   & 42    & 6     & 290   & 40    & 15.5  & 2.4   \\
\vzy          & 31    & 6     & 3.7   & 0.8   & 35    & 7     & 123   & 25    & 2.7   & 0.5   \\
\Zy           & 119   & 7     & 10.7  & 1.2   & 23390 & 180   & 29300 & 800   & 7.7   & 0.7   \\
\tty          & 422   & 34    & 1530  & 90    & 38.9  & 3.3   & 418   & 33    & 31.8  & 2.9   \\
\jfake        & 89    & 34    & 160   & 90    & 220   & 80    & 15100 & 800   & 5.5   & 1.6   \\
\efake        & 34    & 5     & 8.1   & 1.3   & 4.4   & 0.8   & 330   & 50    & 80    & 13    \\
\midrule
Total         & 941   & 29    & 1730  & 40    & 23730 & 160   & 45560 & 230   & 143   & 12    \\
\midrule
Data          & \multicolumn{2}{c}{957} & \multicolumn{2}{c}{1724} & \multicolumn{2}{c}{23707} & \multicolumn{2}{c}{45570} & \multicolumn{2}{c}{146} \\

\bottomrule
\end{tabular}
\end{table}

To obtain the \wwy fiducial production cross-section, a fiducial volume is defined at particle level
to closely match the selection criteria of the SR. Events are selected if they contain at least one photon, and one muon and one electron with opposite electric charges.
Electrons and muons, excluding those originating from $\tau$-lepton decays, are \enquote{dressed} by adding all nearby photons within a cone of size $\Delta R = 0.1$. The dressed leptons are then required to have $\pT > \SI{20}{\GeV}$ and $|\eta| < 2.5$. Final-state photons must be prompt, with $\pT > \SI{20}{\GeV}$ and $\abseta < 2.37$, and satisfy a particle-level isolation requirement:\
the scalar sum of the transverse energy ($E_\mathrm{T}$) of all stable particles, excluding neutrinos, within $\Delta R = 0.2$ of the photon must be less than 7\% of the photon $E_\mathrm{T}$, i.e.\ $E^{\text{iso}}_{\mathrm{T}}/E^{\gamma}_{\mathrm{T}} < 0.07$.
Photons within $\Delta R = 0.4$ of a selected muon or electron are discarded. Additionally, at least one muon or electron must have $\pT > \SI{27}{\GeV}$.
Finally, events with particle-level jets containing $b$-hadrons are removed, where jets are required to have $\pT > \SI{20}{\GeV}$ and $\abseta < 4.5$.
This requirement rejects about $1\%$ of events when evaluated on the \wwy samples.
The fiducial cross-section includes non-resonant and off-shell \wwy contributions, which amount to ${<}1\%$ of the predicted fiducial cross-section.
Using the signal MC sample described in \Cref{sec:data_and_mc}, the predicted fiducial cross-section is $\sigma_{\text{fid}} = 6.1^{\,+1.0}_{-0.7}\,\text{fb}$.
Renormalisation and factorisation scale uncertainties contribute $^{\,+17}_{-11}\%$, evaluated from the envelope of seven-point scale variations where the scales are raised or lowered by a factor of two, excluding combinations where one scale is raised and the other lowered.
The PDF uncertainty contributes 0.13\%, evaluated using the RMS of the 100 replicas in the \NNPDF[3.0nnlo] PDF set.  An uncertainty of 0.8\% arises from varying the strong coupling constant \alphas by $\pm 0.001$ via the PDFs named NNPDF30\_nnlo\_as\_119 and NNPDF30\_nnlo\_as\_117.

The measured \wwy fiducial production cross-section is $6.2 \pm 0.8\,\text{(stat.)} \pm 0.6\,\text{(sys.)}\,\si{fb} = 6.2 \pm 1.0\,\si{fb}$, which corresponds to a total uncertainty of 16\%. The total systematic uncertainty is calculated by combining all nuisance parameter impacts in quadrature. The impact of a nuisance parameter on the signal strength is quantified by the corresponding off-diagonal element of the post-fit covariance matrix~\cite{Pinto:2023yob}. In the case of nuisance parameters representing the MC statistical uncertainty in each bin, their impacts are evaluated by dividing their post-fit off-diagonal elements by their corresponding pre-fit uncertainties. The uncertainty of the measured cross-section is dominated by a statistical uncertainty of 12\%, calculated as the square root of the difference of the squares of the total uncertainty and the total systematic uncertainty.
The impact of systematic uncertainties on the measured signal strength is grouped into categories and summarised in~\Cref{tab:fit_unblined_uncertainty}.

\begin{table}[h!t]
\centering
\caption{The impact of systematic uncertainties on the measured \wwy fiducial cross-section, decomposed into their major categories. For each category, the contribution is obtained by summing in quadrature the impacts of its nuisance parameters. The total systematic uncertainty is calculated similarly by combining all nuisance parameter impacts in quadrature. The statistical uncertainty is calculated as the square root of the difference of the squares of the total uncertainty and the total systematic uncertainty.}
\label{tab:fit_unblined_uncertainty}
\begin{tabular}{lc}
\toprule
Uncertainty source &  $\Delta \sigma / \sigma$ [\%]  \\
\toprule

Non-prompt-photon background modelling  & 5.4 \\
Jet energy scale and resolution & 4.1 \\
Jet flavour tagging & 3.0 \\
MC statistics & 2.9 \\
Signal modelling & 2.8 \\
\Pileup modelling & 2.5 \\
Lepton reconstruction and calibration & 2.3 \\
Prompt-photon background modelling & 2.1 \\
Photon reconstruction and calibration  & 2.0 \\
\MET reconstruction and calibration & 1.1 \\
Luminosity & 0.9 \\
\midrule
\textbf{Total systematic uncertainty} & 9.6 \\
\midrule
\textbf{Statistical uncertainty} & 12~~ \\
\midrule
\textbf{Total uncertainty} & 16~~~~\; \\

\midrule
\end{tabular}
\end{table}

The probability of the background processes alone producing a signal-like excess at least as large as that seen in data is derived using the profile-likelihood-ratio test statistic in the asymptotic approximation~\cite{Cowan:2010js}. The significance of the observed signal is $5.9\sigma$, compared to an expected significance of $6.0\sigma$.


%

%
\section{EFT interpretation}
\label{sec:EFT}


Potential deviations from the SM predictions, such as in cross-section values or kinematic distributions, are modelled using an EFT framework, which extends the SM Lagrangian to include higher-dimensional operators that parameterise new physics at energy scales beyond the SM. In this analysis, the Eboli parameterisation~\cite{Eboli_EFT} is used, focusing exclusively on dimension-8 operators acting on quartic gauge-boson couplings. Dimension-6 operators relevant to triple gauge-boson couplings have already been tightly constrained by ATLAS~\cite{STDM-2016-15,STDM-2017-03,STDM-2017-18,STDM-2017-24} and CMS~\cite{CMS-SMP-16-017,CMS-SMP-16-018,CMS-SMP-17-011,CMS-SMP-18-002,CMS-SMP-18-004,CMS-SMP-18-008,CMS-SMP-19-001} analyses of diboson processes.
Since the \wwy channel offers substantially weaker sensitivity compared with existing diboson measurements, the contributions of dimension-6 operators are not investigated in this analysis.

The $\PW\PW\gamma\gamma$ and $\PW\PW\PZ\gamma$ vertices are sensitive to 13 dimension-8 operators defined in Ref.~\cite{Eboli_EFT}, called \EFTOperator{M0}, \EFTOperator{M1}, \EFTOperator{M2}, \EFTOperator{M3}, \EFTOperator{M4}, \EFTOperator{M5}, \EFTOperator{M7}, \EFTOperator{T0}, \EFTOperator{T1}, \EFTOperator{T2}, \EFTOperator{T5}, \EFTOperator{T6}, and \EFTOperator{T7}.
Each operator is associated with a Wilson coefficient ($f_i$), which parametrises the strength of the corresponding higher-dimensional interaction.
These operators correspond to two classes of mixed-scalar operators, consisting of two covariant derivatives of the Higgs field and two field-strength tensors, and tensor-type operators, consisting of four field-strength tensors.

The effects of higher-dimensional beyond-the-SM (BSM) operators are more pronounced at high photon $\pT$. To enhance the sensitivity of the analysis to such effects, the SR used for the SM \wwy measurement is split into two regions: a one-bin high-$\pT(\gamma)$ signal region with $\pT(\gamma) > \SI{500}{\GeV}$, and a CR with $\pT(\gamma) < \SI{500}{\GeV}$, where the BDT distribution is used. The threshold at \SI{500}{\GeV} is chosen to optimise the expected signal-to-background ratio for BSM contributions, resulting in negligible BSM sensitivity in all CRs including the low-$\pT(\gamma)$ CR, while maximising it in the high-$\pT(\gamma)$ bin. The photon $\pT$ distribution used to define these regions is provided in the Appendix~\ref{app:appendix}.
The expected yield, including the SM \wwy contribution, in the high-$\pT(\gamma)$ region is $2.24 \pm 0.31$, whereas one event is observed in data.

A simultaneous binned maximum-likelihood fit is performed for each dimension-8 operator independently, neglecting possible interference between operators, and using the high-$\pT(\gamma)$ EFT region together with the CRs, following the same set-up as in the \wwy measurement. The SM background normalisations, including that of SM \wwy, are treated as free parameters in the likelihood fit. Additional unconstrained fit parameter is defined such that $f_i^2$ is the normalisation factor of events arising from BSM contributions, while events arising from interference between SM and BSM amplitudes are scaled by $f_i$. Additional modelling uncertainties for the BSM contributions are taken into account, including those from the PDF, \muR and \muF. A one-dimensional 95\% confidence interval (CI) for each Wilson coefficient is derived from the profile-likelihood curves using the criterion $- 2\cdot\Delta\ln\mathcal{L} = 3.84$, as prescribed by Wilks' theorem~\cite{Wilks:1938dza}.

The expected and observed 95\% CIs for the Wilson coefficients are presented in \Cref{tab:eft_limits}. The sensitivity is statistically limited:\ including all systematic uncertainties changes the expected interval widths by less than 1\%.
These limits are comparable to limits obtained from vector-boson scattering measurements and $Z(\nu\nu)\gamma$~\cite{STDM-2020-14}, $\PW\gamma jj$~\cite{STDM-2018-31} and $VVZ$ productions~\cite{STDM-2020-08}.

\begin{table}[!ht]
\centering
\caption{Expected and observed non-unitarised 95\% CIs for the 13 dimension-8 Wilson coefficients.
The expected CIs are derived from an Asimov dataset constructed in two steps:\ first, the control regions, including the SM \wwy CR, are fitted assuming no EFT contribution to extract the background normalisations; second, a signal-plus-background fit is performed using an Asimov dataset built from these fitted normalisations.}
\label{tab:eft_limits}
\begin{tabular}{c|c|c}
\hline
EFT operator  & Expected 95\% CI [\WilsonCoeffUnit{}] & Observed 95\% CI [\WilsonCoeffUnit{}] \\
\hline
\WilsonCoeff{M0} & $[-8, 8]$ & $[-6, 6]$ \\
\WilsonCoeff{M1} & $[-13, 13]$ & $[-10, 10]$ \\
\WilsonCoeff{M2} & $[-3.1, 3.1]$ & $[-2.5, 2.5]$ \\
\WilsonCoeff{M3} & $[-5, 5]$ & $[-4, 4]$ \\
\WilsonCoeff{M4} & $[-8, 8]$ & $[-6, 6]$ \\
\WilsonCoeff{M5} & $[-6, 6]$ & $[-5, 5]$ \\
\WilsonCoeff{M7} & $[-26, 26]$ & $[-21, 21]$ \\
\WilsonCoeff{T0} & $[-1.4, 1.4]$ & $[-1.1, 1.1]$ \\
\WilsonCoeff{T1} & $[-1.7, 1.7]$ & $[-1.4, 1.3]$ \\
\WilsonCoeff{T2} & $[-4, 4]$ & $[-3.1, 3.0]$ \\
\WilsonCoeff{T5} & $[-1.1, 1.1]$ & $[-0.9, 0.9]$ \\
\WilsonCoeff{T6} & $[-1.3, 1.3]$ & $[-1.1, 1.1]$ \\
\WilsonCoeff{T7} & $[-3.0, 3.0]$ & $[-2.4, 2.4]$ \\
\hline
\end{tabular}
\end{table}

Limits on the dimension-8 Wilson coefficients are also evaluated as functions of a cut-off scale, $\sqrt{\hat{s}}$, using the clipping technique~\cite{Brivio:2022pyi}. For events with $\sqrt{\hat{s}}$ greater than the cut-off, the EFT contribution is set to zero. The unitarity bounds for aQGC operators, as derived in Ref.~\cite{Almeida:2020ylr}, are formulated for 2-to-2 scattering processes in proton–proton collisions. To extend these bounds to \wwy production, $\sqrt{\hat{s}}$ is set to the maximum of the three diboson invariant mass combinations, $\max(m_{V_{i}V_{j}})$.

\begin{figure}[!ht]
\centering
\includegraphics[width=.4\textwidth]{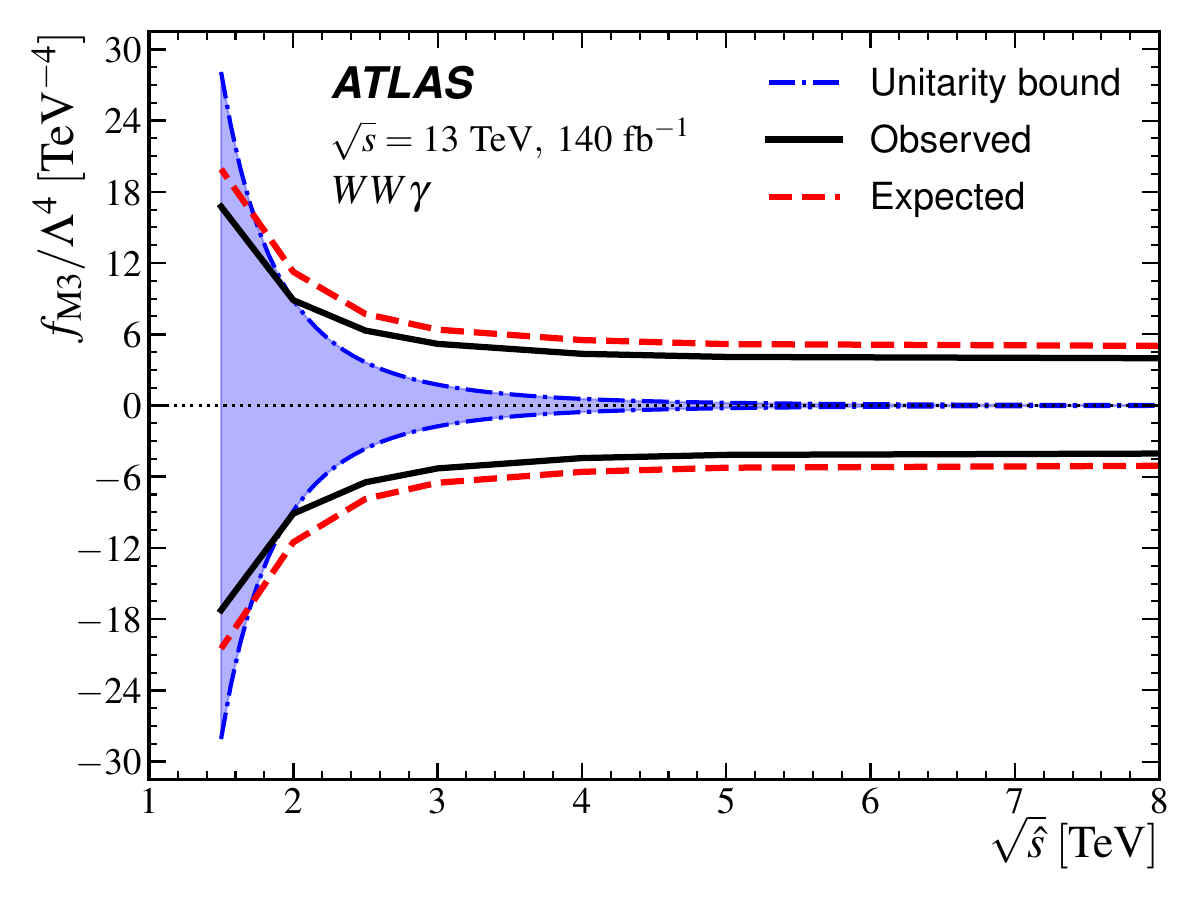}
\caption{
The 95\% confidence interval of the \WilsonCoeff{M3} Wilson coefficient as a function of the energy threshold used in the clipping method to restore unitarity. The blue dash-dotted lines indicate the unitarity bound. The black solid lines show the observed upper and lower limits, while the red dashed lines show the expected upper and lower limits.}
\label{fig:unitarity_clipping_M0}
\end{figure}

The most stringent limit respecting unitarity is found at the intersection of the calculated limits with the unitarity bounds. Among the 13 coefficients investigated, four coefficients exhibit a range of values that respect unitarity: $[-9.5, 9.5]$ \WilsonCoeffUnit at $\sqrt{\hat{s}}=\SI{1.9}{\TeV}$ for \WilsonCoeff{M3}, $[-10, 10]$ \WilsonCoeffUnit at \SI{2.0}{\TeV} for \WilsonCoeff{M5}, $[-4, 4]$ \WilsonCoeffUnit{} at \SI{1.5}{\TeV} for \WilsonCoeff{T6}, and $[-10.5, 10.5]$~\WilsonCoeffUnit at \SI{1.5}{\TeV} for \WilsonCoeff{T7}. \Cref{fig:unitarity_clipping_M0} shows the 95\% CI for the \WilsonCoeff{M3} Wilson coefficient as a function of the cut-off scale applied to restore unitarity.

%

%
%
%

%
\FloatBarrier
\section{Conclusion}
%


The production of \wwy events is observed with a significance of $5.9\sigma$, compared to an expected significance of $6.0\sigma$, using the \emuy final state and \lumi of data from $ \SI{13}{\TeV}$ proton--proton collisions recorded by the ATLAS detector at the LHC between 2015 and 2018.

The measured fiducial cross-section in the \emuy final state is $6.2 \pm 0.8\,\text{(stat.)} \pm 0.6\,\text{(sys.)}~\si{fb} = 6.2 \pm 1.0~\si{fb}$, corresponding to a total uncertainty of 16\%. This measurement is in good agreement with the SM  prediction of $\SI[parse-numbers=false]{6.1^{\,+1.0}_{-0.7}}{\fb}$, calculated at NLO in QCD for $\wwy$\,+\,0-jet events and at LO for events with one or two additional jets.

In addition, constraints at the 95\% confidence level are set on 13 dimension-8 EFT Wilson coefficients within the effective field theory framework. These constitute the first limits from the \wwy channel, probing aQGC at high photon transverse momentum. They are complementary to existing multiboson results and provide valuable input to the global EFT interpretation of QGC.


%

%
%
\section*{Acknowledgements}
%

%
%

%
%

We thank CERN for the very successful operation of the LHC and its injectors, as well as the support staff at
CERN and at our institutions worldwide without whom ATLAS could not be operated efficiently.

The crucial computing support from all WLCG partners is acknowledged gratefully, in particular from CERN, the ATLAS Tier-1 facilities at TRIUMF/SFU (Canada), NDGF (Denmark, Norway, Sweden), CC-IN2P3 (France), KIT/GridKA (Germany), INFN-CNAF (Italy), NL-T1 (Netherlands), PIC (Spain), RAL (UK) and BNL (USA), the Tier-2 facilities worldwide and large non-WLCG resource providers. Major contributors of computing resources are listed in Ref.~\cite{ATL-SOFT-PUB-2025-001}.

We gratefully acknowledge the support of ANPCyT, Argentina; YerPhI, Armenia; ARC, Australia; BMWFW and FWF, Austria; ANAS, Azerbaijan; CNPq and FAPESP, Brazil; NSERC, NRC and CFI, Canada; CERN; ANID, Chile; CAS, MOST and NSFC, China; Minciencias, Colombia; MEYS CR, Czech Republic; DNRF and DNSRC, Denmark; IN2P3-CNRS and CEA-DRF/IRFU, France; SRNSFG, Georgia; BMFTR, HGF and MPG, Germany; GSRI, Greece; RGC and Hong Kong SAR, China; ICHEP and Academy of Sciences and Humanities, Israel; INFN, Italy; MEXT and JSPS, Japan; CNRST, Morocco; NWO, Netherlands; RCN, Norway; MNiSW, Poland; FCT, Portugal; MNE/IFA, Romania; MSTDI, Serbia; MSSR, Slovakia; ARIS and MVZI, Slovenia; DSI/NRF, South Africa; MICIU/AEI, Spain; SRC and Wallenberg Foundation, Sweden; SERI, SNSF and Cantons of Bern and Geneva, Switzerland; NSTC, Taipei; TENMAK, T\"urkiye; STFC/UKRI, United Kingdom; DOE and NSF, United States of America.

Individual groups and members have received support from BCKDF, CANARIE, CRC and DRAC, Canada; CERN-CZ, FORTE and PRIMUS, Czech Republic; COST, ERC, ERDF, Horizon 2020, ICSC-NextGenerationEU and Marie Sk{\l}odowska-Curie Actions, European Union; Investissements d'Avenir Labex, Investissements d'Avenir Idex and ANR, France; DFG and AvH Foundation, Germany; Herakleitos, Thales and Aristeia programmes co-financed by EU-ESF and the Greek NSRF, Greece; BSF-NSF and MINERVA, Israel; NCN and NAWA, Poland; La Caixa Banking Foundation, CERCA Programme Generalitat de Catalunya and PROMETEO and GenT Programmes Generalitat Valenciana, Spain; G\"{o}ran Gustafssons Stiftelse, Sweden; The Royal Society and Leverhulme Trust, United Kingdom.

In addition, individual members wish to acknowledge support from CERN: European Organization for Nuclear Research (CERN DOCT); Chile: Agencia Nacional de Investigaci\'on y Desarrollo (FONDECYT 1230812, FONDECYT 1240864, Fondecyt 3240661, Fondecyt Regular 1240721); China: Chinese Ministry of Science and Technology (MOST-2023YFA1605700, MOST-2023YFA1609300), National Natural Science Foundation of China (NSFC - 12175119, NSFC 12275265); Czech Republic: Czech Science Foundation (GACR - 24-11373S), Ministry of Education Youth and Sports (ERC-CZ-LL2327, FORTE CZ.02.01.01/00/22\_008/0004632), PRIMUS Research Programme (PRIMUS/21/SCI/017); EU: H2020 European Research Council (ERC - 101002463); European Union: European Research Council (BARD No. 101116429, ERC - 948254, ERC 101089007), European Regional Development Fund (SMASH COFUND 101081355, SLO ERDF), European Union, Future Artificial Intelligence Research (FAIR-NextGenerationEU PE00000013), Italian Center for High Performance Computing, Big Data and Quantum Computing (ICSC, NextGenerationEU); France: Agence Nationale de la Recherche (ANR-21-CE31-0022, ANR-22-EDIR-0002, ANR-24-CE31-0504-01); Germany: Baden-Württemberg Stiftung (BW Stiftung-Postdoc Eliteprogramme), Deutsche Forschungsgemeinschaft (DFG - 469666862, DFG - CR 312/5-2); China: Research Grants Council (GRF); Italy: Istituto Nazionale di Fisica Nucleare (ICSC, NextGenerationEU), Ministero dell'Università e della Ricerca (NextGenEU 153D23001490006 M4C2.1.1, NextGenEU I53D23000820006 M4C2.1.1, NextGenEU I53D23001490006 M4C2.1.1, SOE2024\_0000023); Japan: Japan Society for the Promotion of Science (JSPS KAKENHI JP22H01227, JSPS KAKENHI JP22H04944, JSPS KAKENHI JP22KK0227, JSPS KAKENHI JP24K23939, JSPS KAKENHI JP24KK0251, JSPS KAKENHI JP25H00650, JSPS KAKENHI JP25H01291, JSPS KAKENHI JP25K01023); Norway: Research Council of Norway (RCN-314472); Poland: Ministry of Science and Higher Education (IDUB AGH, POB8, D4 no 9722), Polish National Science Centre (NCN 2021/42/E/ST2/00350, NCN OPUS 2023/51/B/ST2/02507, NCN UMO-2019/34/E/ST2/00393, UMO-2022/47/O/ST2/00148, UMO-2023/49/B/ST2/04085, UMO-2023/51/B/ST2/00920, UMO-2024/53/N/ST2/00869); Portugal: Foundation for Science and Technology (FCT); Spain: Ministry of Science and Innovation (MCIN \& NextGenEU PCI2022-135018-2, MICIN \& FEDER PID2021-125273NB, RYC2019-028510-I, RYC2020-030254-I, RYC2021-031273-I, RYC2022-038164-I), Ministerio de Ciencia, Innovación y Universidades/Agencia Estatal de Investigaci\'on (PID2022-142604OB-C22); Sweden: Carl Trygger Foundation (Carl Trygger Foundation CTS 22:2312), Swedish Research Council (Swedish Research Council 2023-04654, VR 2021-03651, VR 2022-03845, VR 2022-04683, VR 2023-03403, VR 2024-05451), Knut and Alice Wallenberg Foundation (KAW 2018.0458, KAW 2022.0358, KAW 2023.0366); Switzerland: Swiss National Science Foundation (SNSF - PCEFP2\_194658); United Kingdom: Royal Society (NIF-R1-231091); United States of America: U.S. Department of Energy (ECA DE-AC02-76SF00515), Neubauer Family Foundation.

%
%


\clearpage
\appendix
\addcontentsline{toc}{part}{Appendices}
\section{Appendix}
\label{app:appendix}


\begin{figure}[!ht]
\centering
\includegraphics[width=.4\textwidth]{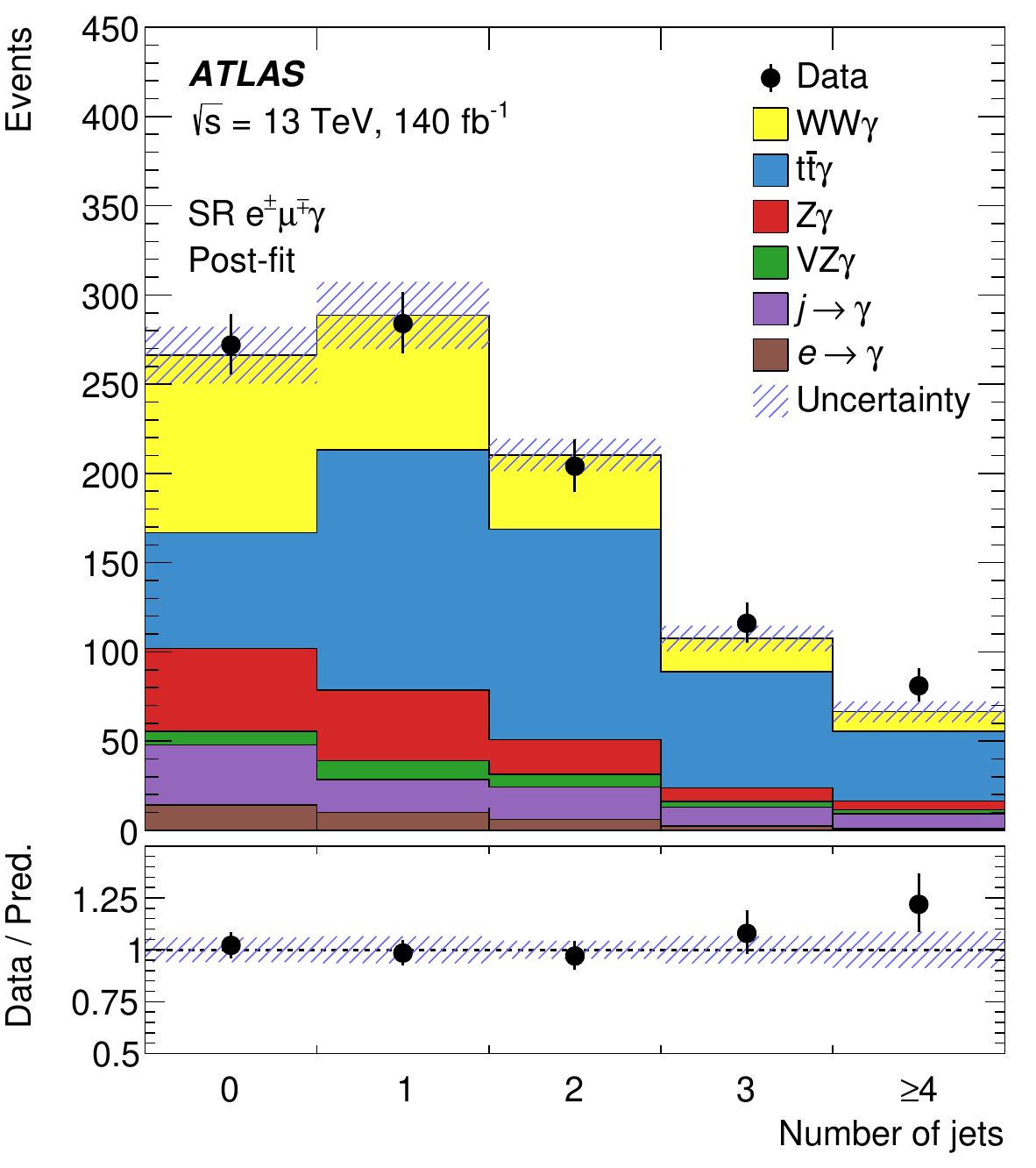}
\caption{
Comparison between data and prediction (\enquote{Pred.}) for the jet multiplicity distribution after the fit to data (\enquote{Post-Fit}) under the signal-plus-background hypothesis, shown in the SR. The uncertainty band includes both statistical and systematic uncertainties as obtained from the fit.
The rightmost bin includes overflow events.  }

\end{figure}

\begin{figure}[!ht]
\centering
\includegraphics[width=.4\textwidth]{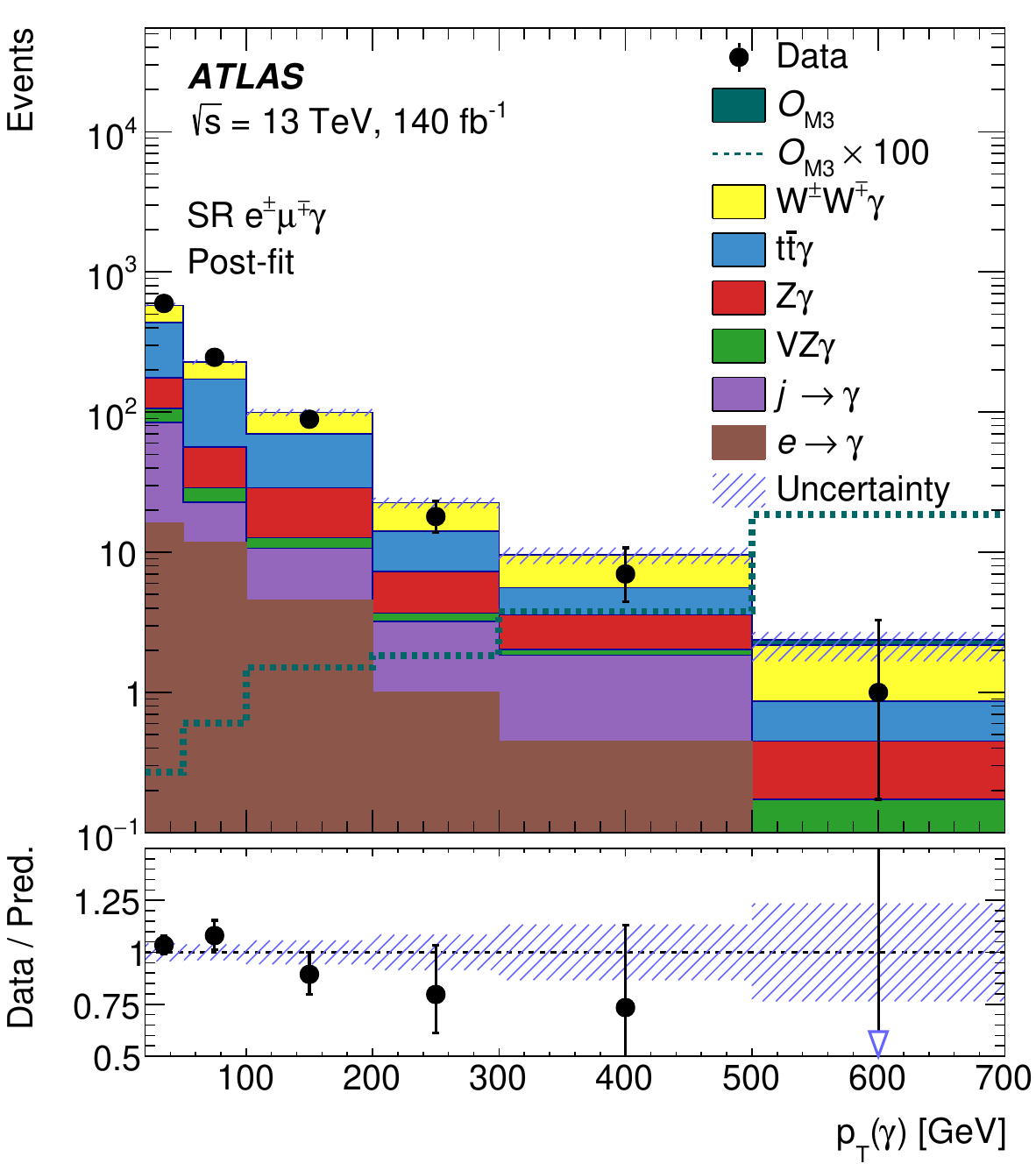}
\caption{
Comparison between data and prediction (\enquote{Pred.}) for the photon \pT distribution after the fit to data (\enquote{Post-fit}) under the signal-plus-background hypothesis including \EFTOperator{M3} operator, shown in the \wwy SR. The uncertainty band includes both statistical and systematic uncertainties as obtained from the fit.
The rightmost bin includes overflow events.
The contribution of the Wilson coefficient \WilsonCoeff{M3} is shown as shaded teal area and is normalised to the prediction for a coefficient value of $1 ~\WilsonCoeffUnit$. The dashed teal line corresponds to the same prediction multiplied by a factor of 100. The EFT sensitivity is enhanced when the photon \pT is above \SI{500}{\GeV}.
}

\end{figure}


\printbibliography

\clearpage
\input{atlas_authlist}
%

%
%

%

%
%
%
%

\end{document}

%% file: atlas_authlist.tex
 
\begin{flushleft}
\hypersetup{urlcolor=black}
{\Large The ATLAS Collaboration}

\bigskip

\AtlasOrcid[0000-0002-6665-4934]{G.~Aad}$^\textrm{\scriptsize 104}$,
\AtlasOrcid[0000-0001-7616-1554]{E.~Aakvaag}$^\textrm{\scriptsize 17}$,
\AtlasOrcid[0000-0002-5888-2734]{B.~Abbott}$^\textrm{\scriptsize 123}$,
\AtlasOrcid[0000-0002-0287-5869]{S.~Abdelhameed}$^\textrm{\scriptsize 119a}$,
\AtlasOrcid[0000-0002-1002-1652]{K.~Abeling}$^\textrm{\scriptsize 55}$,
\AtlasOrcid[0000-0001-5763-2760]{N.J.~Abicht}$^\textrm{\scriptsize 49}$,
\AtlasOrcid[0000-0002-8496-9294]{S.H.~Abidi}$^\textrm{\scriptsize 30}$,
\AtlasOrcid[0009-0003-6578-220X]{M.~Aboelela}$^\textrm{\scriptsize 45}$,
\AtlasOrcid[0000-0002-9987-2292]{A.~Aboulhorma}$^\textrm{\scriptsize 36e}$,
\AtlasOrcid[0000-0001-5329-6640]{H.~Abramowicz}$^\textrm{\scriptsize 157}$
% ...
\bigskip
\\

% ...
$^{17}$Department for Physics and Technology, University of Bergen, Bergen; Norway.\\
% ...
$^{30}$Physics Department, Brookhaven National Laboratory, Upton NY; United States of America.\\
% ...
$^{36}$$^{(a)}$Facult\'e des Sciences Ain Chock, Universit\'e Hassan II de Casablanca;$^{(b)}$Facult\'{e} des Sciences, Universit\'{e} Ibn-Tofail, K\'{e}nitra;$^{(c)}$Facult\'e des Sciences Semlalia, Universit\'e Cadi Ayyad, LPHEA-Marrakech;$^{(d)}$LPMR, Facult\'e des Sciences, Universit\'e Mohamed Premier, Oujda;$^{(e)}$Facult\'e des sciences, Universit\'e Mohammed V, Rabat;$^{(f)}$Institute of Applied Physics, Mohammed VI Polytechnic University, Ben Guerir; Morocco.\\
% ...
$^{45}$Physics Department, Southern Methodist University, Dallas TX; United States of America.\\
% ...
$^{49}$Fakult\"{a}t Physik, Technische Universit{\"a}t Dortmund, Dortmund; Germany.\\
% ...
$^{55}$II. Physikalisches Institut, Georg-August-Universit\"{a}t G\"ottingen, G\"ottingen; Germany.\\
% ...
$^{104}$CPPM, Aix-Marseille Universit\'e, CNRS/IN2P3, Marseille; France.\\
% ...
$^{119}$$^{(a)}$New York University Abu Dhabi, Abu Dhabi;$^{(b)}$United Arab Emirates University, Al Ain; United Arab Emirates.\\
% ...
$^{123}$Homer L. Dodge Department of Physics and Astronomy, University of Oklahoma, Norman OK; United States of America.\\
% ...
$^{157}$Raymond and Beverly Sackler School of Physics and Astronomy, Tel Aviv University, Tel Aviv; Israel.\\
% ...
$^{a}$ Also at Affiliated with an institute formerly covered by a cooperation agreement with CERN.\\
$^{b}$ Also at An-Najah National University, Nablus; Palestine.\\
$^{c}$ Also at Borough of Manhattan Community College, City University of New York, New York NY; United States of America.\\
$^{d}$ Also at Center for Interdisciplinary Research and Innovation (CIRI-AUTH), Thessaloniki; Greece.\\
$^{e}$ Also at Centre of Physics of the Universities of Minho and Porto (CF-UM-UP); Portugal.\\
$^{f}$ Also at CERN, Geneva; Switzerland.\\
% ...

\end{flushleft}
